\documentclass[journal,12pt,onecolumn,draftclsnofoot,]{IEEEtran}
\usepackage{cite}
\usepackage{amsmath,amssymb,amsfonts}
\usepackage{algorithmic}
\usepackage{graphicx}
\usepackage{textcomp}
\def\BibTeX{{\rm B\kern-.05em{\sc i\kern-.025em b}\kern-.08em
    T\kern-.1667em\lower.7ex\hbox{E}\kern-.125emX}}

\usepackage{xcolor,soul,framed} 
\usepackage{cite}
\usepackage{amsmath,amssymb,amsfonts}
\usepackage{caption} 
\usepackage{subfig}
\usepackage{braket}
\usepackage{algorithmic}
\usepackage{graphicx}
\usepackage{textcomp}
\usepackage{xspace}
\usepackage{amsthm}
\usepackage{hyperref}
\usepackage{gensymb}
\usepackage[nolist]{acronym}
\usepackage{titlesec}
\titlespacing*{\subsection}{0pt}{-1pt}{1pt}
\theoremstyle{plain}

\usepackage{subfig}
\usepackage{tabularx}
\usepackage{booktabs} 
\colorlet{shadecolor}{yellow}
\usepackage{array}

\begin{document}
\bstctlcite{}
    \title{A Hybrid Noise Approach to Modelling of Free-Space Satellite Quantum Communication Channel for Continuous-Variable QKD}
  \author{Mouli~Chakraborty,~\IEEEmembership{Student Member,~IEEE,}
      Anshu~Mukherjee,~\IEEEmembership{Member,~IEEE,}\\
      Ioannis~Krikidis,~\IEEEmembership{Fellow,~IEEE,}
      Avishek~Nag,~\IEEEmembership{Senior Member,~IEEE,}
      and~Subhash~Chandra,~\IEEEmembership{Member,~IEEE}

  \thanks{This publication has emanated from research supported in part by a grant from Science Foundation Ireland under Grant number 18/CRT/6222. For the purpose of Open Access, the author has applied a CC BY public copyright licence to any Author Accepted Manuscript version arising from this submission. This publication has emerged from research supported partly by a grant from IEEE Antennas and Propagation Society Graduate Fellowship Program - Quantum Technologies Initiative under Grant number `IEEE Funding – 1421.9040008''}
  \thanks{M. Chakraborty and S. Chandra are with the School of Natural Science, Trinity College Dublin, The University of Dublin, College Green, Dublin 02, Ireland (e-mail: chakrabm@tcd.ie) Ireland (e-mail: chakrabm@tcd.ie; SCHANDRA@tcd.ie).}
  \thanks{A. Mukherjee is with the School of Electrical and Electronic Engineering, University College Dublin, Belfield, Dublin 04, Ireland (e-mail: anshu.mukherjee@ieee.org).}%
  \thanks{I. Krikidis is with the IRIDA Research Centre for Communication Technologies, Department of Electrical and Computer Engineering, University of Cyprus, 1678 Nicosia, Cyprus (e-mail: krikidis@ucy.ac.cy).}
  \thanks{A. Nag is with the School of Computer Science, University College Dublin, Belfield, Dublin 04, Ireland (e-mail: avishek.nag@ucd.ie).}
  }


\maketitle

\begin{abstract}

This paper significantly advances the application of \ac{QKD} in \ac{FSO} satellite-based quantum communication. We propose an innovative satellite quantum channel model and derive the secret quantum key distribution rate achievable through this channel. Unlike existing models that approximate the noise in quantum channels as merely Gaussian distributed, our model incorporates a hybrid noise analysis, accounting for both quantum Poissonian noise and classical \ac{AWGN}. This hybrid approach acknowledges the dual vulnerability of \ac{CV} Gaussian quantum channels to both quantum and classical noise, thereby offering a more realistic assessment of the quantum \ac{SKR}. This paper delves into the variation of \ac{SKR} with the \ac{SNR} under various influencing parameters. We identify and analyze critical factors such as reconciliation efficiency, transmission coefficient, transmission efficiency, the quantum Poissonian noise parameter, and the satellite altitude. These parameters are pivotal in determining the \ac{SKR} in \ac{FSO} satellite quantum channels, highlighting the challenges of satellite-based quantum communication. Our work provides a comprehensive framework for understanding and optimizing \ac{SKR} in satellite-based \ac{QKD} systems, paving the way for more efficient and secure quantum communication networks.

\end{abstract}

\begin{IEEEkeywords}Free-space optical channel, Continuous variables Gaussian quantum channel, Satellite quantum communication, Hybrid quantum noise, Quantum key distribution, Quantum secret key rate.
\end{IEEEkeywords}

\section{Introduction}
\label{sec:introduction}

Quantum communication signifies a paradigm shift in information transmission, leveraging the unique attributes of quantum mechanics to enable novel communication methods such as quantum teleportation and entanglement swapping \cite{usenko2012entanglementQKD}. Central to these innovations is quantum teleportation, which allows for the disassembly and remote reconstruction of delicate quantum states, facilitating the instantaneous transport of quantum information over vast distances \cite{vallone2015experimentalQSatComm}. Recent milestones in this field include the deterministic teleportation of photonic quantum bits across tens of kilometers and the broadcast of quantum states through multinode networks, thereby opening new avenues for intricate quantum operations \cite{mastriani2021satellite}. Beyond the mere transfer of information, quantum communication ensures enhanced security and inviolability, paving the way for ultrasecure communication channels.

The integration of quantum communication with satellite technology has opened new frontiers in global secure communications, addressing the distance constraints inherent in terrestrial quantum networks \cite{bedington2017SAT_QKD}. Satellite-based quantum communication systems leverage the vast reach of orbiting platforms to establish quantum links across continental scales, potentially enabling a global quantum internet \cite{khatri2021SpookySATQKD}. This synergy between quantum and satellite technologies promises to revolutionize fields such as cryptography, distributed computing, and precision timekeeping on a global scale. Central to the advancement of quantum communication is the development of Quantum Key Distribution (\ac{QKD}), a cryptographic protocol that enables the secure exchange of encryption keys between distant parties \cite{gisin2002quantum}. Utilizing the unique properties of quantum states, Gaussian continuous variables (\ac{CV}) \ac{QKD} allows two parties to establish secret keys with security guaranteed by the laws of quantum physics rather than computational complexity \cite{fossier2009GasussianCVQKD},\cite{bedington2017SAT_QKD}. Despite its potential, implementing \ac{QKD} over extended ranges and in noisy environments remains challenging, necessitating constant innovation and refinement to address real-world communication scenarios.

One of the promising approaches to overcoming these challenges involves space-based \ac{QKD} solutions that employ satellites as relay nodes, creating a globally interconnected quantum network \cite{serafini2017quantumCVbook},\cite{lodewyck2007quantum}. The experimental foundations of \ac{QKD} date back to seminal works that introduced protocols such as BB84 and its subsequent enhancements, incorporating decoy states and bit error rate optimizations \cite{liao2017satqkd}. These innovations have improved robustness against \ac{PNS} attacks and increased key generation rates. Space-based experiments have successfully demonstrated free-space BB84 \ac{QKD} between ground stations separated by hundreds of kilometers and quasi-single-photon transmissions from satellites to Earth \cite{moli2009performanceSAT},\cite{sidhu2022finitekeySat}. Recent advancements include the SpeQtral-1 quantum satellite mission, which aims to establish ultra-secure communications via \ac{QKD}, and ongoing work on CubeSats explicitly designed for spaceborne \ac{QKD} protocols and its applications \cite{ahmadi2024},\cite{nguyen2023eSatQKD}.

Despite this progress, significant challenges persist in satellite quantum communications. Optical transmission between satellites and ground-based platforms faces several limitations, including diffraction, systematic pointing errors, and atmospheric turbulence \cite{hughes2000quantumSat},\cite{barry1985beam}. These factors cause transmission losses by broadening the optical beam and introducing wavelength-dependent losses due to atmospheric absorption \cite{giggenbach2022atmospheric}. Detector efficiencies need substantial improvement for efficient quantum key exchange \cite{khmelev2024eurasian}. 
Photon loss during propagation through atmospheric turbulence and background noise mitigation are critical areas of investigation \cite{bourgoin2013SatComm}.  While larger telescopes and shorter photon wavelengths can mitigate diffraction losses, a balance between telescope size, cost, and transmission performance is necessary \cite{kurt2021hapsSat},\cite{khmelev2024eurasian}. Atmospheric turbulence, caused by temperature-induced refractive index fluctuations, broadens and wanders the beam, introducing additional loss \cite{maharjan2022atmosphericTurbuSatComm},\cite{hosseinidehaj2015quantum}. Works like \cite{andrews1995Sat} have developed \acp{p.d.f.} of transmittance for slant propagation paths, proposed models for atmospheric quantum channels with turbulence, and experimentally characterized distant \ac{FSO} atmospheric channels. Addressing these challenges is essential for the seamless integration of \ac{QKD} into existing satellite communication infrastructures, enhancing the robustness and efficiency of global quantum networks with quantum repeaters \cite{mastriani2020satellite}. Adaptive optical systems and optimal ground station selection can mitigate turbulence effects, particularly in uplink channels where turbulence is most significant \cite{dmytryszyn2021lasers_uplink_downlink},\cite{PirandolasatQComm2021}.

This paper addressed the limitations of current \ac{QKD} systems by proposing a novel free-space atmospheric satellite quantum channel model.  The proposed model uniquely incorporates quantum Poissonian noise and classical additive white Gaussian noise (\ac{AWGN}), offering a realistic representation of the noise characteristics impacting quantum channels. It identifies and analyzes critical parameters influencing the \ac{SKR}, including reconciliation efficiency, transmission coefficient, transmission efficiency, satellite altitude, and the quantum Poissonian noise parameter. By examining how \ac{SKR} varies with signal-to-noise ratio (\ac{SNR})  under different conditions, we provide valuable insights into the potential and limitations of satellite-based quantum communication systems. 
The comprehensive analysis of atmospheric and environmental factors and advanced quantum communication protocols underscores quantum satellite communications' robustness and high potential for global-scale secure networks. Our work demonstrates the efficacy of the hybrid quantum noise model in enhancing free-space quantum satellite communication technologies, presenting new solutions to address quantum noise challenges in quantum channels.

This work focuses on quantum key distribution through a quantum channel model for free-space satellite communication. Previous works, such as those by \cite{dequal2021_SAT_CV_QKD}, have proposed atmospheric quantum satellite communication channel models that consider noise as  \ac{AWGN}, typically expected in classical channels \cite{lapidoth2002AWGN}. Building on this, \cite{chakraborty2024hybridquantumnoiseapproximation, mouli2024} introduced a mathematical model incorporating hybrid quantum noise to depict the quantum channel accurately. Our current work extends these models to propose an accurate estimation of the \ac{SKR} for free-space satellite quantum communication. In essence, the critical contributions of this paper can be summarized in three main aspects

\begin{itemize}

    \item   The paper introduces a pioneering free-space atmospheric satellite quantum channel model that integrates both quantum Poissonian noise and classical \ac{AWGN}. This hybrid noise approach is a significant advancement over traditional models that typically consider only Gaussian noise. By acknowledging the dual vulnerabilities of continuous-variable Gaussian quantum channels to both quantum and classical noise, the proposed model offers an accurate and realistic representation satellite-based \ac{FSO} quantum channel's capacity. This innovation enhances the accuracy of channel's capacity, making it a crucial contribution to satellite-based quantum communication.\\

    \item   The paper makes a substantial contribution to the practical \ac{QKD} rates in satellite-based communication.  he paper provides valuable insights into optimizing quantum communication systems by exploring how the \ac{SKR} varies with the \ac{SNR} under different conditions.  The work not only advances theoretical models but also provides a robust framework for the practical design and implementation of \ac{QKD} systems, making it an exemplary study in quantum communications.\\

     \item The paper thoroughly examines different factors like reconciliation efficiency, electronic noises, detection efficiency, excess noise, transmission coefficient, transmission efficiency, satellite altitude, and the quantum Poissonian noise parameter. It excels in identifying and analyzing key parameters that influence the \ac{SKR} in free-space atmospheric satellite quantum channels. This finding underscores the challenges of satellite-based quantum communication and the need for careful optimization of system parameters.

\end{itemize}


Our comprehensive approach and findings pave the way for enhanced design and implementation of robust, efficient, and secure global quantum communication networks. 

\textit{Notation:} We use $tr$ denote the trace of a matrix, and $\mathbf{T(\cdot)}$ represent a trace-preserving map $\mathbf{T}$. The adjoint of a matrix $\mathbf{A}$ is denoted by $\mathbf{A}^\dag$, while its transpose is represented as $\mathbf{A}^t$. The complex conjugate of a vector $\nu$ is written as $\nu^*$. The symbol $\otimes$ is used to denote the tensor product, and $\mathbb{C}$ represents the set of complex numbers. The parameter $\lambda$ indicates the Poisson parameter, and $\mathcal{N}$ is used for Gaussian density. Quantum states are represented using Dirac notation, where $\ket{\cdot}$ denotes a ket and $\bra{\cdot}$ denotes a bra. The notation $|\cdot|$ represents the norm. Reconciliation efficiency is denoted by $\beta$, while $\nu_{ele}$ represents electronic noise. The parameter $\eta$ stands for detection efficiency, and $\epsilon$ denotes excess noise. The transmission coefficient is represented by $T$, and the transmission efficiency by $\tau$. Lastly, the symbol $h$ denotes the altitude of the satellite.

\section{System Model for Free-space Satellite Quantum Communication}

\subsection{The Uplink and Downlink Scenarios in Quantum Satellite Communication }\vspace{0.15cm}

\begin{figure}[!t]
\centering
\includegraphics[width=0.75\columnwidth]{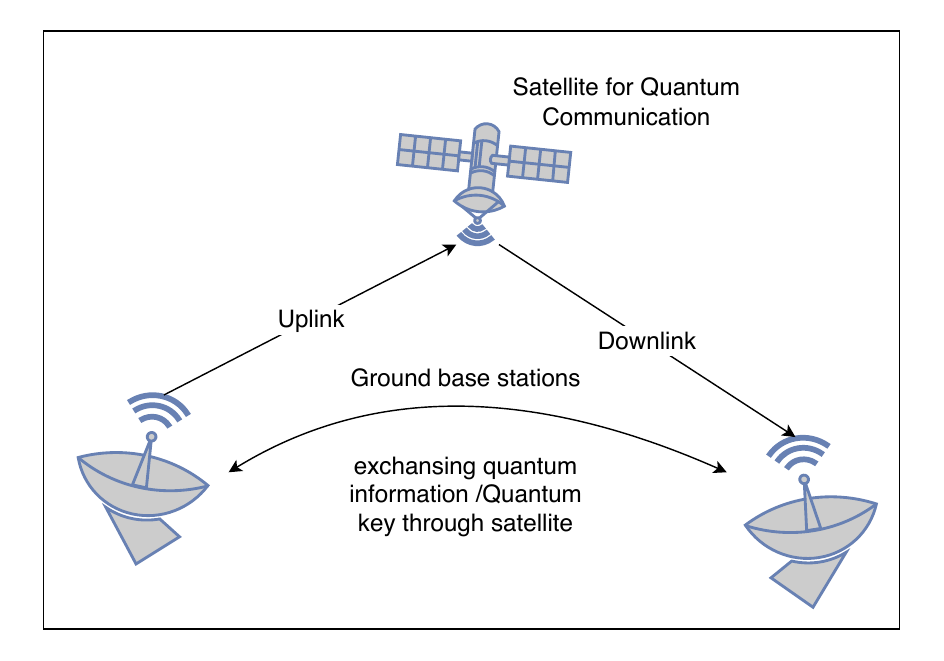}
\caption{A basic model for Quantum Satellite Communication, can be 
replicable with multiple base stations or satellites for feasibility of long-distance communication and quantum key distribution, it also forms the basis of the quantum internet. } 
\label{fig basic_model_for_Qsat_comm}
\end{figure}

In satellite-based quantum communication, there are two primary channels: uplink (ground-to-satellite) and downlink (satellite-to-ground). In the uplink channel, the ground station transmits signals to the satellite receiver, while in the downlink channel, the satellite transmits signals to the ground station receiver. This is shown in Fig.~\ref{fig basic_model_for_Qsat_comm}. In this setup, complex quantum engineering components are confined to the ground stations, where one station acts as the source of quantum states and the other as the receiver. Although reflecting quantum states via a satellite is a sophisticated engineering task, it avoids the need for onboard generation of quantum communication information. Deploying quantum technology at the ground stations offers practical advantages such as lower-cost maintenance and rapid upgrades as new technology matures. Several possible architectures and schemes for implementing satellite-based quantum communication depend on the types of links utilized, as illustrated and studied in the existing literature in terms of \ac{CV}-\ac{QKD}. For example, some schemes \cite{HanzoSatcomm2019} involve higher complexity and deployment of quantum technology on the satellite. Some schemes \cite{HanzoSatcomm2019} have the source of quantum states onboard the satellite, with both ground stations acting as receivers. In another example, a scheme \cite{HanzoSatcomm2019} involves both ground stations transmitting quantum states to the satellite, where quantum measurements are performed on the received states. The classical measurement results are then communicated back to the ground stations. These measurement results support entanglement swapping and \ac{MDI} protocols, facilitating \ac{QKD} between the two ground stations.
\vspace{0.15cm}

\subsection{A Generalised CV-Gaussian Quantum Channel Model }\vspace{0.15cm}

\begin{figure*}
	\centerline{\includegraphics[width = 1\textwidth]{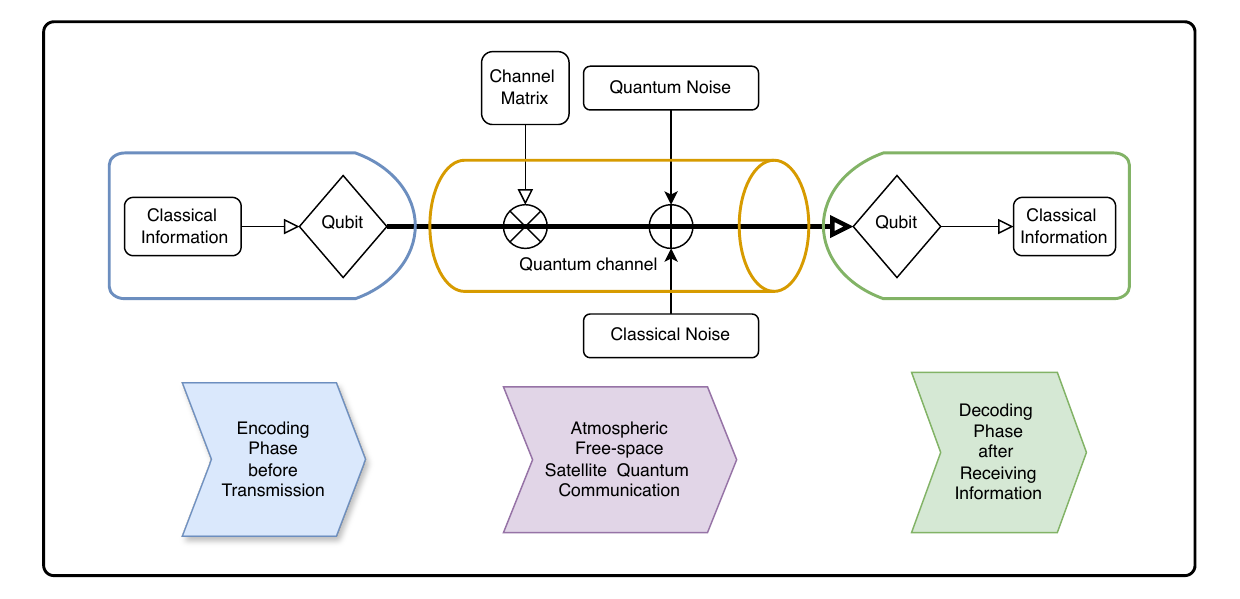}}
	\caption{ A schematic diagram showing a single quantum link between transmitter and receiver.}
	\label{quantum noisy channel diag}
\end{figure*}

The \ac{CV} Gaussian quantum channel model used in this work is adapted from our previous study \cite{chakraborty2024hybridquantumnoiseapproximation, mouli2024},\cite{Mouli2024MLQComm}, presented in this section. For additional insights, we request the readers to consult the comprehensive overview provided in our prior work. Quantum communication involves the transfer of quantum states via a quantum channel and consists of three main steps: (I) preparation of quantum states where classical information is encoded into quantum states; (II) transmission of these quantum states over a quantum channel, such as optical fiber or free-space optical \ac{FSO} channel; and (III) detection where the received states are decoded using quantum measurement to yield classical information. This configuration is shown in Fig.~\ref{quantum noisy channel diag}. Quantum states in communication are represented using \ac{DV} and \ac{CV} descriptions. \ac{DV} technology encodes information into discrete features, such as photon polarization, using single-photon detectors \cite{HanzoSatcomm2019}.

A qubit, the fundamental unit of \ac{DV} quantum information, is a superposition of two orthogonal quantum states: \(\ket{\boldsymbol{\psi}} = c_1 \ket{\boldsymbol{0}} + c_2 \ket{\boldsymbol{1}}\), where \(c_1\) and \(c_2\) are complex coefficients satisfying \(|c_1|^2 + |c_2|^2 = 1\). In contrast, \ac{CV} encoding maps information onto the optical field's quadrature variables, spanning an infinite-dimensional Hilbert space. Detection of \ac{CV} states often involves homodyne or heterodyne detectors, which are faster and more efficient than single-photon detectors. Quantum states, whether pure or mixed, are described by density matrices. For a pure state qubit \(\ket{\boldsymbol{\psi}}\), the density matrix is \(\boldsymbol{\rho} = \ket{\boldsymbol{\psi}} \bra{\boldsymbol{\psi}}\), while mixed states are represented as probabilistic mixtures of pure states \cite{demoen1977completely, holevo1999capacity}.

A quantum channel is defined as a \ac{CPTP} map, which transforms density operators \(\boldsymbol{\rho}\) within a Hilbert space \(\mathcal{H}\). This transformation can be described as \(\boldsymbol{\rho } \mapsto \mathbf{T}(\boldsymbol{\rho})\). For any quantum channel \(\mathbf{T}\), there exists a corresponding state \(\boldsymbol{\rho}_E\) in an environment Hilbert space \(\mathcal{H}_E\) and a unitary operation \(\mathbf{U}\), satisfying \(\mathbf{T}(\boldsymbol{\rho}) = \text{tr}_E[\mathbf{U}(\boldsymbol{\rho} \otimes \boldsymbol{\rho}_E)\mathbf{U}^\dagger]\), where the environment \(E\) drives the decoherence process \cite{holevo2001evaluating, lindblad2000cloning}. In Gaussian channels, the transformation can be expressed as \(\boldsymbol{\rho} \mapsto \mathbf{A}^t \boldsymbol{\rho } \mathbf{A} + \mathbf{Z}\), where \(\mathbf{A}\) modulates the signal, and \(\mathbf{Z}\) accounts for both quantum and classical noise. In special cases when \(\mathbf{A = I}_{2n}\), where $n$ is the degrees of freedom of the corresponding quantum system, the channel can be simplified to \(\boldsymbol{\rho} \mapsto \boldsymbol{\rho} + \mathbf{Z}\), and the transformation remains valid as a quantum channel if sufficient noise is introduced. A qubit, represented as a point vector \((\theta, \phi, r)\) on the Bloch sphere, can be treated as a scalar variable \(\theta\) under hybrid noise constraints \cite{mouli2024, Mouli2024MLQComm, Mouli2024MECOM}. Under this scalar transformation, the scalar variable representing the qubit can be mapped in the following way $\theta \mapsto \theta  +Z $ where $Z$ is the random variable representing the hybrid quantum noise $Z$ consists of quantum Poisson noise and classical \ac{AWGN}. Hence, the generalized quantum channel equation can be written from the above as \(Y = X + Z\) where \(X\) and  \(Y\) are the transmitted signal and the received signal, respectively \cite{mouli2024, Mouli2024MLQComm}. The channel equation has been derived in the following section for the special case of the satellite quantum channel. \vspace{0.15cm}

\subsection{The CV-Gaussian Quantum Channel Model for Satellite Quantum Communication }
\vspace{0.15cm}
 
This paper examines a quantum satellite communication model comprising a ground station and a satellite, enabling bidirectional transmission. We designate the sender as Alice and the receiver as Bob, regardless of location, and include an eavesdropper, Eve, who may intercept transmissions anywhere along the path. Alice prepares quantum states, typically qubits or \ac{CV} states encoded as photons, which are transmitted via a satellite-based free-space optical channel.  This channel is subject to free-space diffraction, atmospheric turbulence, and thermal noise, affecting signal quality differently in uplink and downlink scenarios.

The satellite serves as an intermediary relay between Alice and Bob. Upon receiving the quantum states, Bob may employ homodyne, heterodyne, or single-photon detection methods depending on the nature of the transmitted quantum states. Bob's ability to accurately decode the information depends on the fidelity of the received quantum states, which is influenced by the noise and losses introduced by the satellite channel. Quantum mechanical principles constrain an eavesdropper's (Eve) potential interception attempts, though she may use sophisticated strategies to extract information covertly. This model allows for comprehensive analysis of quantum communication protocols and security scenarios in satellite-based systems, addressing the unique challenges of long-distance, free-space quantum transmission under various environmental conditions and potential security threats.

From this generalized description of a quantum channel, we can drive towards the atmospheric \ac{FSO} satellite quantum channel model represented by the following equation \cite{dequal2021_SAT_CV_QKD}

\begin{equation}
    Y=TX +Z,
    \label{eq atmospheric quantum channel}
\end{equation} 
where the quantum channel transmitted signal is given by random variables $X$, representing Alice's inputs, and the received signal $Y$ at Bob's side, where $T$ is the overall transmission coefficient for the single use of the quantum channel. The focus should be on the transmission coefficient $T$ (with the transmission efficiency $\tau= T^2$). The goal is to maximize the \ac{SKR} that Alice and Bob can extract while minimizing the information that Eve can gain. This model captures the complexity of satellite quantum communication and underscores the importance of accounting for various noise sources and channel imperfections in ensuring secure communication.

\begin{figure}%
    \centering
    \subfloat[\centering ]{{\includegraphics[width=7.5cm]{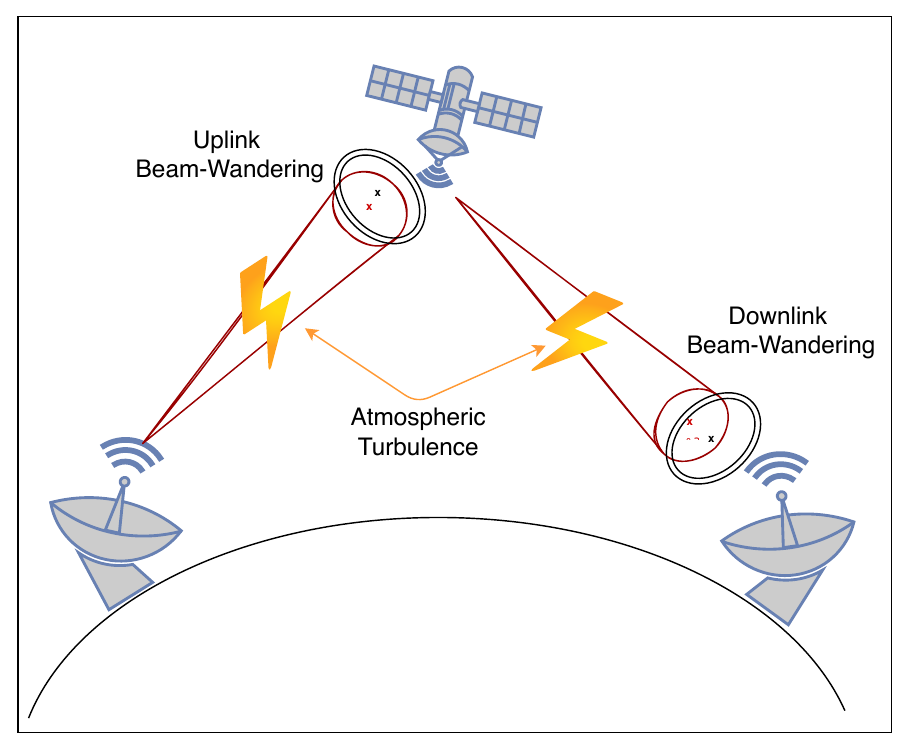} }}%
    \qquad
    \subfloat[\centering ]{{\includegraphics[width=7.5cm]{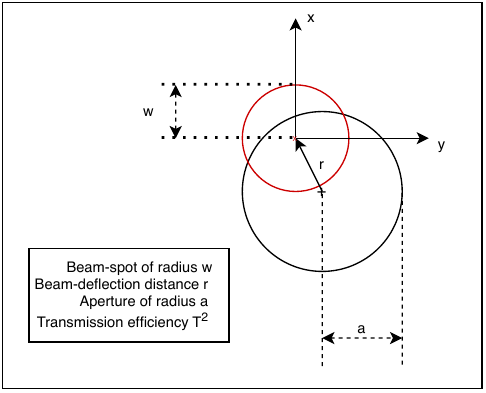} }}%
    \caption{(a) The atmospheric turbulence with beam-wandering. Due to various atmospheric conditions, considering the beam-wandering model for uplinks and downlinks for satellites is useful, and it shows realistic scenarios in the detection of the secret key rate. (b) The beam-wandering model for laser direction. The red circle shows the position of the incoming beam on the receiver's aperture; the black circle represents the receiver's aperture. }%
    \label{fig atmosphericturbulence_with_beamwandering}%
\end{figure}

In \ac{FSO}, the primary sources of loss include diffraction, absorption, scattering, and atmospheric turbulence. Under favorable weather conditions, diffraction-induced beam spreading and turbulence-induced beam wandering are the primary challenges, while absorption, scattering, and scintillation have less impact. Diffraction causes the light beam to spread, leading to divergence losses, particularly if the receiver's aperture is too small to capture the entire beam. Mitigating these losses can involve increasing the receiver's aperture or using shorter wavelengths, though this requires balancing cost and efficiency. Narrower beams reduce diffraction losses but are more prone to alignment errors. Absorption and scattering, caused by atmospheric particles, depend on the wavelength and can be minimized by selecting appropriate communication wavelengths. These effects are often ignored in theoretical models for quantum communication, but adverse weather conditions like fog, rain, or snow can significantly reduce channel transmissivity, posing challenges to FSO communication in such environments \cite{HanzoSatcomm2019}. These scenarios are pictorially shown in Fig.~\ref{fig atmosphericturbulence_with_beamwandering} (a).


\section{Methodology of Quantum Cryptography for Free-space Satellite Quantum Communication}

\subsection{Secret Quantum Key Distribution through \ac{FSO} Satellite Quantum Channel}\vspace{0.15cm}


This section outlines the methodology for calculating the key generation rate across a fading channel. To initiate the process, Alice generates a sequence of \(2N\) real random variables \(X_{1}, X_{2}, \cdots, X_{2N}\) that are Gaussian distributed with variance \(\sigma_{X_k}^2\), expressed as
$X_{k} \sim \mathcal{N}(x;\mu_{X_k}, \sigma_{X_k}^2)$ ,
with the associated \ac{p.d.f.} given by \cite{dequal2021_SAT_CV_QKD}

\begin{equation}
  f_{X_k}(x) = \frac{1}{\sigma_{X_k}\sqrt{2\pi}} e^{-\frac{1}{2}\left(\frac{x - \mu_{X_k}}{\sigma_{X_k}}\right)^2},
  \label{eq The p.d.f. of transmitted signal}
\end{equation}
where \(\mu_{X_k}\) and \(\sigma_{X_k}\) are the mean and standard deviation (s.d.) of the distribution, respectively. These \(2N\) random variables \(X_{1}, X_{2}, \cdots, X_{2N}\) correspond to \(N\) coherent states given by \(\ket{\boldsymbol{a}_{1}}, \ket{\boldsymbol{a}_{2}}, \cdots, \ket{\boldsymbol{a}_{N}}\), where \(\boldsymbol{a}_k = X_{2k-1} + iX_{2k} \in \mathbb{C}\) and \(i = \sqrt{-1}\). Each state is sent through the quantum channel to Bob, who performs heterodyne detection, simultaneously measuring both quadratures. For the \(k\)th use of the channel, Bob obtains two results, \(Y_{2k-1}\) and \(Y_{2k}\), corresponding to \(X_{2k-1}\) and \(X_{2k}\), respectively. The raw key string \(\boldsymbol{Y} = (Y_{1}, Y_{2}, \cdots, Y_{2N})\) is obtained, which is particularly useful under low transmission efficiency with reverse reconciliation. The asymptotic values are considered in the limiting case \(N \rightarrow \infty\). The formula traditionally used to calculate the asymptotic value of the \ac{SKR}, particularly under reverse reconciliation, is known as the Devetak-Winter bound \cite{dequal2021_SAT_CV_QKD,ghalaii2022quantumSatcomm}

\begin{equation}
  K_{WD} = \beta I_{AB} - \chi_{BE},
  \label{eq DW bound}
\end{equation}
where \(\beta I_{AB}\) measures the correlation between Alice and Bob's datasets, incorporating the inefficiency of the error correction process through the parameter \(\beta\) (\( \leq 1\)). Meanwhile, \(\chi_{BE}\) indicates the amount of information Eve has about the raw key derived from Bob's data. 


\subsection{Propagation Channel Model for \ac{FSO} Satellite Quantum Communication}
\vspace{0.15cm}


To assess the efficacy of a \ac{QKD} protocol over a specific quantum channel, one must calculate the mutual information \(\beta I_{AB}\) and the Holevo quantity \(\chi_{BE}\). These quantities are essential for determining the \ac{SKR}. 
The parameter estimation process must be modeled to predict the expected outcomes for Alice and Bob within the given channel model. The quantum channel connecting Alice and Bob can be represented as a phase-insensitive noisy Bosonic channel. This formulation requires that the random variables \(X_k\)  and \(Y_k\)  are considered within this framework. The relationship between \(X_k\) and \(Y_k\) is given by 

\begin{equation}
    Y_k = T_k X_k + Z_k,
    \label{eq quantum channel}
\end{equation}
where \(T_k\) is the overall transmission coefficient for the \(k\)-th channel use, with the transmission efficiency of $k$-th channel \(\tau_k= T_k^2 \) , and \(Z_k\) represents the hybrid quantum noise. The distribution of the hybrid quantum noise \(Z_k\) is given by \cite{mouli2024}

\begin{equation} 
  \begin{split}
   f_{Z_{k}}(z) & = \sum_{j=0}^{\infty} \frac{e^{-\lambda}\lambda^j}{j!}\frac{1}{\sigma_{Z_{k}^{(2)}}\sqrt{2\pi}} e^{-\frac{1}{2}\left(\frac{z-j-\mu_{Z_{k}^{(2)}}}{\sigma_{Z_{k}^{(2)}}}\right)^2}\\
       & = \sum_{j=0}^{\infty} u_{j}^{(Z_{k})}\mathcal{N} \left(z; \mu_{j}^{(Z_{k})},\, {\sigma_{j}^{(Z_{k})}}^{2}\right),
    \label{eq p.d.f. of hybrid quantum noise}
  \end{split}
\end{equation} 
where \(u_{j}^{(Z_{k})} = \frac{e^{-\lambda}\lambda^j}{j!}\), \(\sum_{j=0}^{\infty} u_{j}^{(Z_{k})} = 1\), and \(u_{j} \geq 0 \quad  \forall j\). The term \(\mathcal{N} (z; \mu_{j}^{(Z_k)},\, {\sigma_{j}^{(Z_k)}}^{2})\) represents a Gaussian density with mean \(\mu_{j}^{(Z_{k})} = \mu_{Z_{k}^{(2)}} + j\), variance \({\sigma_{j}^{(Z_k)}}^{2} = \sigma_{Z_{k}^{(2)}}^2\) and standard deviation \({\sigma_{j}^{(Z_k)}}\). The hybrid quantum noise \(Z_k\) is a combination of quantum Poissonian noise \(Z_{k}^{(1)}\) and classical additive white Gaussian noise \(Z_{k}^{(2)}\), such that \(Z_{k} = Z_{k}^{(1)} + Z_{k}^{(2)}\). Quantum Poissonian Noise \(Z_{k}^{(1)}\) 
follows the \ac{p.m.f.} given by $f_{Z_{k}}^{(1)} (j) = \frac{e^{-\lambda}\lambda^j}{j!}$, where \(\lambda \geq 0\), \(\lambda \in \{0, 1, 2, \ldots\}\), and \(j \in \{0, 1, 2, \ldots\}\). The classical \ac{AWGN}  \(Z_{k}^{(2)}\) follows Gaussian distribution with the mean \(\mu_{Z_{k}^{(2)}}\) and variance \(\sigma_{Z_{k}^{(2)}}^{2}\). \vspace{0.15cm}
 

\subsection{Analyzing the Transmission Efficiency for Satellite Quantum Channel}
\vspace{0.15cm}

This section delves into the critical aspects of transmission efficiency in free-space quantum communication channels, specifically focusing on satellite-based systems. The transmission efficiency \(\tau_k\) is characterized by a random variable that ranges between 0 and 1. The probability distribution of \(\tau_k\) aligns with the log-normal distribution, approximating the \ac{PDTC} in entanglement distribution scenarios. The probability distribution of \(\tau_k\) is given by \cite{erven2012free-spaceQKD, ghalaii2022quantumSatcomm,vasylyev2012SemenovVogel,wang2018pdtc_freespaceQKD}
\begin{equation}
\begin{split}
\mathcal{P}(\tau_k) & = \frac{1}{\sqrt{2\pi}\sigma \tau_k} e^{-\frac{1}{2}\left(\frac{\ln\tau_k +\ln\Tilde{\tau}_k}{\sigma}\right)^2} .
\label{eq PDTC}
\end{split}
\end{equation}
Here, \(\tau_k\) represents the atmospheric transmittance, \(-\ln\Tilde{\tau}_k\) is the logarithm of the mean atmospheric transmittance, and \(\sigma\) is the variance of \(-\ln\tau_k\), which characterizes the atmospheric turbulence. The primary effects of atmospheric turbulence on quantum communication include beam spreading, scintillation, and beam wandering \cite{galetsky2022leoSatellite, HanzoSatcomm2019}: (I) Beam Spreading occurs when the beam width is significantly larger than the eddies' radius in the turbulent layer. (II) Scintillation dominates when the beam width and eddies' radius are comparable, which is especially significant for satellite downlink. (III) Beam Wandering is common in both uplinks and downlinks. It involves time-dependent random lateral shifts in the beam's position. The beam wandering effect, shown in Fig.~\ref{fig atmosphericturbulence_with_beamwandering} (b), is influenced by how atmospheric turbulence causes lateral shifts in the beam's position.

The Weibull distribution is employed to measure the pointing error of the \ac{S/C} which is given by the p.d.f.
\(P(r,\sigma_{r})= \frac{r}{\sigma_r^2} e^{-\left(\frac{r}{\sqrt{2}\sigma_r}\right)^2},\)
where \(\sigma_r\) represents the standard deviation, and \(r\) is the displacement distance from the receiver's center. The \ac{PDTC} statistical model represents the transmission efficiency \(T\) received at the \ac{GS} \cite{vasylyev2012SemenovVogel, HanzoSatcomm2019}
$T^2 = T_0^2 e^{-\left(\frac{r}{R_1}\right)^{\kappa_1}}$,
where $T_0^2 = 1 - e^{-2\frac{a^2}{W^2}}$, and \(\kappa_1\) and \(R_1\) are the shape and scale parameters given by 
\begin{equation}
    \kappa_1 = 8\frac{a^2}{W^2} \frac{e^{-4\frac{a^2}{W^2}} I_1\left(4\frac{a^2}{W^2}\right) }{ 1 - e^{-\frac{4a^2}{W^2}}I_0\left(4\frac{a^2}{W^2}\right)}\left[\ln\left(\frac{2T_0^2}{1 - e^{-4\frac{a^2}{W^2}}I_0\left(\frac{4a^2}{W^2}\right)}\right)\right]^{-1}
\end{equation}
and \begin{equation}
    R_1 = a\left[\ln\left(\frac{2T_0^2}{1 - e^{-4\frac{a^2}{W^2}}I_0\left(4\frac{a^2}{W^2}\right)}\right)\right]^{-1/\kappa_1},
\end{equation} 
where \(a\) is the receiver aperture radius, \(W\) is the beam size, and \(I_0\) and \(I_1\) are the modified Bessel functions of the first kind.

Considering the \ac{PDTC} and beam wandering, we account for the \ac{S/C}'s off-axis pointing behavior. Assuming the beam fluctuates around the aperture's center, we can calculate the \ac{PDTC} (\(P\)) as follows  \cite{vasylyev2012SemenovVogel}
\begin{equation}
    P(T) = \frac{2R_1^2}{\sigma_r^{\kappa_1} T} \left(2\ln\left(\frac{T_0}{T}\right)\right)^{\frac{2}{\kappa_1}-1} e^{-\frac{1}{2\sigma_r R_1^2}\left(2\ln\left(\frac{T_0}{T}\right)\right)^{\frac{2}{\kappa_1}}}.
\end{equation}

\ac{PDTC} is crucial for long-distance quantum communication, which relies on entanglement. 
This model, focused on beam wandering, underpins expansions into quantum free-space channels, suggesting that fluctuating losses may better conserve quantum properties like entanglement and squeezing than constant losses. This preservation is vital for successful free-space quantum communication.\vspace{0.15cm}

\subsection{Received Signal Model for \ac{FSO} Satellite Quantum Channel}
\vspace{0.15cm}

In quantum communication, accurately calculating the key rate over fading channels is essential, especially in free-space satellite communication, where atmospheric turbulence significantly influences transmission efficiency. 
Considering the impact of atmospheric turbulence, the transmission coefficient \( T_k \) varies but is assumed stable over short periods. The transmission coefficient remains stable during back-to-back uses, allowing Alice and Bob to adjust error correction strategies, assuming approximate knowledge of \(T_k\). Therefore, it is worth considering a fixed value of $T_k$ during the consecutive transmission through the quantum channel and writing $T$ instead of $T_k$, where $T$ is a constant value of transmission coefficient, and the modified channel equation becomes \cite{dequal2021_SAT_CV_QKD,PirandolasatQComm2021}
\begin{equation}
    Y_k = T X_k + Z_k.
    \label{eq quantum channel}
\end{equation}
For given transmitted signal \( X_k \sim \mathcal{N}(x;\mu_{X_k}, \sigma_{X_k}^2) \), the scaled signal \( T X_k \sim \mathcal{N}(s;T \mu_{X_k}, T^2 \sigma_{X_k}^2) \). The \ac{p.d.f.} of \( T X_k \) is given by
\begin{equation}
  f_{TX_k}(s) = \frac{1}{T \sigma_{X_k} \sqrt{2\pi}} e^{-\frac{1}{2} \left( \frac{s - T \mu_{X_k}}{T \sigma_{X_k}} \right)^2}.
  \label{eq The p.d.f. of TX_k}
\end{equation}

The hybrid noise \( Z_k \) is given in \eqref{eq p.d.f. of hybrid quantum noise}, hence the \ac{p.d.f.} of the received signal \( Y_k \) is obtained in convolution of \( f_{TX_k} \) and \( f_{Z_k} \) given by 
\begin{multline}
f_{Y_k}(y)
        =\sum_{j=0}^{\infty}
             \frac{e^{-\lambda}\lambda^j}{j!}\frac{1}{\sqrt{2\pi(\sigma_{X_k}^2 +T^2\sigma_{Z_{j}^{(2)}}^2)}} e^{\left[-\frac{\big(y-j-\mu_{X_k}-T\mu_{Z_{k}^{(2)}}\big)^2}{2\big(\sigma_{X_k}^2 +T^2\sigma_{Z_{j}^{(2)}}^2\big)}\right]}\\
        =\sum_{j=0}^{\infty}v_{j}^{(Y_{k})}\mathcal{N} \Big(y;\mu_{j}^{(Y_{k})}, \,{\sigma_{j}^{(Y_{k})}}^{2}\Big),
  \label{eq final p.d.f. of Y2} 
\end{multline}
where $v_{j}^{(Y_{k})}=\frac{e^{-\lambda}\lambda^j}{j!}$ , $\sum_{j=0}^{\infty} v_{j}^{(Y_{k})}=1$, $v_{j}^{(Y_{k})}\geq 0 \quad  \forall j \in \{0, 1, 2, \cdots \}$ and 
$\mathcal{N}(y;\mu_{j}^{(Y_{k})},\,{\sigma_{j}^{(Y_{k})}}^{2})$ is Gaussian density in dummy variable $y$, with mean $\mu_{j}^{(Y_{k})}=T\mu_{X_k}+\mu_{Z_{k}^{(2)}}+j$,  s.d. 
$\sigma_{j}^{(Y_{k})}=\sqrt{(T^2\sigma_{X_k}^2 +\sigma_{Z_{k}^{(2)}}^2)}$ and variance ${\sigma_{j}^{(Y_{k})}}^{2}$.

By assuming a stable transmission coefficient \( T \) over short periods and analyzing the hybrid quantum noise, we derive the \ac{p.d.f.} of the received signal in free-space quantum communication channels. This allows us to understand the key generation rate and the impact of various noise components, ultimately aiding in designing efficient quantum key distribution protocols.

\vspace{0.15cm}
\subsection{Approximation of The Quantum Signal Models for Entropy Estimation}
\vspace{0.15cm}

This section analyzes the channel's entropy by approximating the hybrid quantum noise and received signal as finite \ac{GMs}. In the next section, we derive upper and lower entropy bounds, offering a robust framework for the satellite quantum channel's capacity. 

The probability distribution of the hybrid quantum noise \(Z_k\) is given by in \eqref{eq p.d.f. of hybrid quantum noise}, the approximation of $f_Z(z)$ is taken as \cite{mouli2024}
\begin{equation} 
  \begin{split}
   f_{Z_{k}}(z) & = \sum_{j=0}^{R} \frac{e^{-\lambda}\lambda^j}{j!}\frac{1}{\sigma_{Z_{k}^{(2)}}\sqrt{2\pi}} e^{-\frac{1}{2}\left(\frac{z-j-\mu_{Z_{k}^{(2)}}}{\sigma_{Z_{k}^{(2)}}}\right)^2}
       = \sum_{j=0}^{R} u_{j}^{(Z_{k})}\mathcal{N} \left(z; \mu_{j}^{(Z_{k})},\, {\sigma_{j}^{(Z_{k})}}^{2}\right),
    \label{eq approx. p.d.f. of hybrid quantum noise}
  \end{split}
\end{equation} 
where \(u_{j}^{(Z_{k})} = \frac{e^{-\lambda}\lambda^j}{j!}\), $\sum_{i=0}^{R} u_{i}^{(Z_{k})} \approx 1$ for large $R$, and \(u_{j}^{(Z_{k})} \geq 0 \quad  \forall j\). The term \(\mathcal{N} (z; \mu_{j}^{(Z_k)},\, {\sigma_{j}^{(Z_k)}}^{2})\) represents a Gaussian density with mean \(\mu_{j}^{(Z_{k})} = \mu_{Z_{k}^{(2)}} + j\) and variance \({\sigma_{j}^{(Z_k)}}^{2} = \sigma_{Z_{k}^{(2)}}^2\), with standard deviation \({\sigma_{j}^{(Z_k)}}\). This is the Gaussian mixture in scalar variable $Z_k$, and the corresponding Gaussian mixture in random vector $\bar{Z}_k \in \mathbb{R}^M$ is 

\begin{equation}
   f_{\bar{Z}_k}(\Bar{z}) = \sum_{i=0}^{R} u_{i}^{(\bar{Z}_k)}\mathcal{N} \Big(\Bar{z};\Bar{\mu_{i}}^{(\bar{Z}_k)},\,{\Sigma}_{i}^{(\bar{Z}_k)}\Big),
    \label{eq approx. p.d.f. of hybrid quantum noise vector}   
\end{equation}
where $\Bar{z}$ is the random vector in $\mathbb{R}^M$, $\Bar{\mu_{i}}^{(\bar{Z}_k)}$ is the mean vector, and ${\Sigma}_{i}^{(\bar{Z}_k)}$ is the covariance matrix of the corresponding Gaussian density $\mathcal{N}$. 


Consider the \ac{p.d.f.} of the received signal $Y_k$ given by in \eqref{eq final p.d.f. of Y2}, the approximation of $f_{Y_k}(y)$ is taken as
\begin{multline}
    f_{Y_{k}}(y) = \sum_{j=0}^{R} 
                   \frac{e^{-\lambda}\lambda^j}{j!}\frac{1}{\sqrt{2\pi(\sigma_{X_{k}}^2 +T^2\sigma_{Z_{k}^{(2)}}^2)}}e^{\left[-\frac{\big(y-j-T\mu_{X_{k}}-\mu_{Z_{k}^{(2)}}\big)^2}{2\big(\sigma_{X_{k}}^2 + T^2\sigma_{Z_{k}^{(2)}}^2\big)}\right]} \\
                 =\sum_{j=0}^{R} 
                  v_{j}^{(Y_{k})}\mathcal{N} \Big(y;\mu_{j}^{(Y_{k})},\,{\sigma_{j}^{(Y_{k})}}^{2}\Big), 
      \label{eq approx. p.d.f. of the received signal}        
\end{multline} 
where \( v_j^{(Y_{k})} = \frac{e^{-\lambda} \lambda^j}{j!} \), $\sum_{i=0}^{R} u_{i}^{(Y_{k})} \approx 1$ for large $R$, \( \mu_j^{(Y_{k})} = T \mu_{X_k} + \mu_{Z_k^{(2)}} + j \), and \( \sigma_j^{(Y_{k})} = \sqrt{T^2 \sigma_{X_k}^2 + \sigma_{Z_k^{(2)}}^2} \). This is the Gaussian mixture in scalar variable $Y_{k}$, and  the corresponding Gaussian mixture in random vector $\Bar{Y}_{k} \in \mathbb{R}^M$ is 

\begin{equation}
   f_{\bar{Y}_k}(\Bar{y}) = \sum_{i=0}^{R} u_{i}^{(\bar{Y}_k)}\mathcal{N} \Big(\Bar{y};\Bar{\mu_{i}}^{(\bar{Y}_k)},\,{\Sigma}_{i}^{(\bar{Y}_k)}\Big), 
    \label{eq approx. p.d.f. of received signal vector}   
\end{equation}
where $\Bar{z}$ is the random vector in $\mathbb{R}^M$, $\Bar{\mu_{i}}^{(\bar{Y}_k)}$ is the mean vector, and ${\Sigma}_{i}^{(\bar{Y}_k)}$ is the covariance matrix of the corresponding Gaussian density $\mathcal{N}$. The detailed calculations and approximation methodology of \acp{p.d.f.} have been showcased in \cite{Mouli2024MLQComm}.

For a continuous-valued random vector \( \bar{x} \in \mathcal{R}^N \) with \ac{p.d.f.} \( f(\bar{x}) \), the differential entropy is \cite{mouli2024}
\begin{equation}
    H(\bar{x}) = E\left[-\log f(\bar{x})\right] = -\int_{\mathcal{R}^N} f(\bar{x}) \log f(\bar{x}) \, d\bar{x}.
    \label{entropy def}
\end{equation}
Now, if p.d.f. $f(\cdot)$ is a  \ac{GMM}, which means it can be represented as a weighted sum of $L$ component Gaussian densities as given by the expression,
 \(f(\bar{x}) = \sum_{i=1}^{L} w_{i} \mathcal{N} \big(\bar{x};\bar{\mu_{i}},\,{\Sigma_{i}}\big),
 \)
where $w_{i}$ are non-negative weighting coefficients with $\sum_{i}^{L} w_{i}=1$ and 
$\mathcal{N}(\bar{x};\bar{\mu_i},{\Sigma_i}) $ is a Gaussian density mean vector $\bar{\mu_i}$, and covariance matrix ${\Sigma_i}$ $\forall i =1(1)L$.
The differential entropy of the hybrid quantum noise \( \bar{Z}_k \) is $H(\bar{Z}_k) = -\int_{\chi_{\bar{Z}_k}} f_{Z_k}(\bar{z}) \log f_{\bar{Z}_k}(\bar{z}) \, d\bar{z}$, and for the received signal \( \bar{Y}_k \), the differential entropy is $H(\bar{Y}_k) = -\int_{\chi_{\bar{Y}_k}} f_{\bar{Y}_k}(\bar{y}) \log f_{\bar{Y}_k}(\bar{y}) \, d\bar{y}$, where \(\chi_{\bar{Z}_k}\) is the support of \( f_{\bar{Z}_k} \) and \(\chi_{\bar{Y}_k}\) is the support of \( f_{\bar{Y}_k} \). The entropy cannot be directly calculated for Gaussian mixtures due to the logarithm of a sum of exponential functions, but it can be approximated using the following bounds. The entropy of Gaussian mixtures, although complex due to the logarithm of a sum of exponentials, can be approximated for practical purposes. In the \ac{QKD} context, accurate entropy calculation is crucial for determining the key rate. Entropy approximation methods often face challenges in quantifying the deviation between actual and approximated values. Therefore, establishing tight lower and upper bounds for the entropy of a Gaussian mixture random vector is essential. These bounds can provide a meaningful approximation and be calculated in the closed form \cite{Huber2008,mouli2024}.

The lower bound \(H_{L}(\Bar{x})\) of the entropy is given by \cite{Huber2008,mouli2024}
\begin{equation}
    H_{L}(\Bar{x})=-\sum_{i=0}^{L}w_{i} \log_{2}\left(\sum_{j=0}^{L}w_{j} m_{i,j}\right),
  \label{eq lower bdd of entropy for Gaussian mixture} 
\end{equation}
where \(m_{i,j} = \mathcal{N} \left(\Bar{\mu_{i}}; \Bar{\mu_{j}},\, \Sigma_{i} + \Sigma_{j}\right)\), \(\bar{\mu_{i}}\) is the mean vector, and \(\Sigma_{i}\) is the covariance matrix of the corresponding component of the Gaussian mixture \(f(\bar{x})\).

The upper bound \(H_{U}(\Bar{x})\) of the entropy is given by \cite{Huber2008,mouli2024}
\begin{equation}
    H_{U}(\Bar{x})=\sum_{i=0}^{L}w_{i} \log_{2}\left(-\log_{2}w_{i}+\frac{1}{2} \log_2{\left((2\pi e)^{N}\left|\Sigma_{i}\right|\right)}\right).
  \label{eq upper bdd of entropy for Gaussian mixture} 
\end{equation}

\subsection{Capacity of \ac{FSO} quantum satellite channel}
\vspace{0.15cm}

Since $f_{\bar{Z}_k}$ and $f_{\bar{Y}_k}$ are both Gaussian mixture densities, the closed-form solution of the corresponding entropies $H_{\bar{Z}_k}$ and $H_{\bar{Y}_k}$ of the noise and the received signal can not be calculated due to the logarithm of the sum and exponential function \cite{Huber2008,mouli2024}. However, \eqref{eq lower bdd of entropy for Gaussian mixture} and \eqref{eq upper bdd of entropy for Gaussian mixture} provide the lower and the upper bounds for each entropy $H_{\bar{Z}_k}$ and $H_{\bar{Y}_k}$.
Let us call the upper and lower bounds of $H_{\bar{Z}_k}$ by $U_{\bar{Z}_k}$ and $L_{\bar{Z}_k}$ and the upper and lower bounds of $H_{\bar{Y}_k}$ by $U_{\bar{Y}_k}$ and $L_{\bar{Y}_k}$, respectively, given by
   \(  L_{\bar{Z}_k} \le H_{\bar{Z}_k} \le U_{\bar{Z}_k} \)
and
  \( L_{\bar{Y}_k} \le H_{\bar{Y}_k} \le U_{\bar{Y}_k}. \)

In \cite{Gyongyosi2018}, the mutual information $I({\Bar{X_{k}}};{\bar{Y}_k})$ between  transmitted signal ${\Bar{X_{k}}}$ and received signal ${\bar{Y}_k}$ is defined as 
 $I({\Bar{X_{k}}};{\bar{Y}_k}) =  H({\Bar{X_{k}}})+H({\bar{Y}_k})-H({\Bar{X_{k}}},{\bar{Y}_k})$
 and applying the chain rule for continuous variables we have,
 $H({\Bar{X_{k}}},{\bar{Y}_k})=H({\bar{Y}_k}|{\Bar{X_{k}}}) -H({\Bar{X_{k}}})$
 Therefore, 
\begin{multline}
I({\Bar{X_{k}}};{\bar{Y}_k}) =  H({\bar{Y}_k})-H({\bar{Y}_k}|{\Bar{X_{k}}}) 
= H({\bar{Y}_k})-H({\Bar{X_{k}}}+{\bar{Z}_k}|{\Bar{X_{k}}})= H({\bar{Y}_k})-H({\bar{Z}_k}|{\Bar{X_{k}}})\\
= H({\bar{Y}_k})-H({\bar{Z}_k})=H({\bar{Y}_k})-H({\bar{Z}_k})
\end{multline}
as transmitted signal ${\Bar{X_{k}}}$ and  noise ${\bar{Z}_k}$ are independent. We rename $H({\bar{Z}_k})$ as $H_{{\bar{Z}_k}}$ for notational easiness, and the bound for mutual information can be calculated as follows
\begin{multline}
-L_{\bar{Z}_k} \geq -H_{\bar{Z}_k} \geq -U_{\bar{Z}_k} 
\implies H_{\bar{Y}_k}-L_{\bar{Z}_k} \geq H_{\bar{Y}_k}-H_{\bar{Z}_k} \geq H_{\bar{Y}_k}-U_{\bar{Z}_k} \\
\implies U_{(\bar{Y}_k)}-L_{(\bar{Z}_k)}\geq H_{(\bar{Y}_k)}-L_{(\bar{Z}_k)} \geq H_{(\bar{Y}_k)}-H_{(\bar{Z}_k)}=I({\Bar{X_{k}}};{\bar{Y}_k}).
\end{multline}   
Hence,
\begin{equation}
 C=\max_{\substack{
         f_{{\bar{X_{k}}}}({\bar{x}})
        }}
        I({\bar{X_{k}}};{\bar{Y}_k}) \leq U_{\bar{Y}_k}-L_{\bar{Z}_k}.
  \label{eq upper bound for capacity for Gaussian input}
\end{equation}
From \eqref{eq lower bdd of entropy for Gaussian mixture}, the lower bound $L_{\bar{Z}_k}$ of entropy $H_{\bar{Z}_k}$, can be calculated as 
\begin{equation}
        L_{\bar{Z}_k}=-\sum_{j=0}^{R}u_{j}^{(\bar{Z}_k)} \log_{2}\Big(\sum_{l=0}^{R}u_{l}^{(\bar{Z}_k)} \mathcal{N} \Big(\Bar{\mu_{j}}^{(\bar{Z}_k)};\Bar{\mu_{l}}^{(\bar{Z}_k)},\,{\Sigma}_{j}^{(\bar{Z}_k)}+{\Sigma}_{l}^{(\bar{Z}_k)}\Big)\Big) .
\end{equation}
From \eqref{eq upper bdd of entropy for Gaussian mixture}, similarly we calculate $U_{\bar{Y}_k}$, the upper bound for $H_{\bar{Y}_k}$, as 
\begin{equation}
   U_{\bar{Y}_k}=\sum_{j=0}^{R}v_{j}^{(\bar{Y}_k)} \log_{2}\Big(-\log_{2}v_{j}^{(\bar{Y}_k)} +\frac{1}{2} \log_2{\Big((2\pi e)^{M}\Big|{\Sigma}_{j}^{(\bar{Y}_k)} \Big|\Big)}\Big),
\end{equation}
where $u_{j}^{(\bar{Z}_k)} =v_{j}^{(\bar{Y}_k)} =\frac{e^{-\lambda}\lambda^j}{j!} \forall j =0(1)R. $ 
Hence, 
\begin{multline}
    U_{(\bar{Y}_k)} -L_{(\bar{Z}_k)} 
=\sum_{j=0}^{R}\frac{e^{-\lambda}\lambda^j}{j!}\Bigg(-\log_{2}\Big(\frac{e^{-\lambda}\lambda^j}{j!}\Big) + \frac{1}{2} \log_2{\Big((2\pi e)^{M}\Big|{\Sigma}_{j}^{(\bar{Y}_k)} \Big|\Big)} \\
+ \log_{2}\Big(\sum_{l=0}^{R}\frac{e^{-\lambda}\lambda^l}{l!} \mathcal{N} 
 \Big(\Bar{\mu_{j}}^{(\bar{Z}_k)} ;\Bar{\mu_{l}}^{(\bar{Z}_k)},\,{\Sigma}_{j}^{(\bar{Z}_k)} +{\Sigma}_{l}^{(\bar{Z}_k)} \Big)\Big)\Bigg). 
 \label{eq difference between upper bdd and lower bdd for Gaussian input}
\end{multline}
From {\eqref{eq upper bound for capacity for Gaussian input}} and \eqref{eq difference between upper bdd and lower bdd for Gaussian input}, the $k$-th quantum channel capacity defined as channel capacity at $k$-th transmission of the key given by
\begin{equation}
   \begin{split}
       C_{k} &
            =\sum_{j=0}^{R}\frac{e^{-\lambda}\lambda^j}{j!}\Bigg(-\log_{2}\Big(\frac{e^{-\lambda}\lambda^j}{j!}\Big)
            + \frac{1}{2} \log_2{\Big((2\pi e)^{M}\Big|{\Sigma}_{j}^{(\bar{Y}_k)} \Big|\Big)}\\
            & +\log_{2}\Big(\sum_{l=0}^{R}\frac{e^{-\lambda}\lambda^l}{l!} \mathcal{N} \Big(\Bar{\mu_{j}}^{(\bar{Z}_k)} ;\Bar{\mu_{l}}^{(\bar{Z}_k)},\,{\Sigma}_{j}^{(\bar{Z}_k)} +{\Sigma}_{l}^{(\bar{Z}_k)} \Big)\Big)\Bigg).
          \label{eq the expression for the channel capacity for Gaussian input vector}
    \end{split}
\end{equation} 
This is the expression for the quantum channel capacity when the noise ${\bar{Z}_k} $ and the received signal ${\bar{Y}_k} $ are random vectors of dimension $M$. In scalar analogy, that is when the noise $Z_{k} $ and the received signal $Y_{k}$ are random variables with \acp{p.d.f.} \eqref{eq p.d.f. of hybrid quantum noise} and \eqref{eq final p.d.f. of Y2} respectively, the expression of the  capacity reduces to
\begin{multline}
C_{k} =\sum_{j=0}^{R}\frac{e^{-\lambda}\lambda^j}{j!}\Bigg(-\log_{2}\Big(\frac{e^{-\lambda}\lambda^j}{j!}\Big)+ \frac{1}{2}\log_2{\Big(2\pi e{\sigma}_{j}^{(Y_k)}\Big)}\\
 +\log_{2}\Big(\sum_{l=0}^{R}\frac{e^{-\lambda}\lambda^l}{l!}\mathcal{N} \Big({\mu_{j}}^{(Z_k)};{\mu_{l}}^{(Z_k)},\,{{\sigma}_{j}^{(Z_k)}}^2 +{{\sigma}_{l}^{(Z_k)}}^2\Big)\Big)\Bigg)
\label{eq the expression for the channel capacity Gaussian input scaler}
\end{multline}      
by putting $M=1$, and replacing $\Big|{\Sigma}_{j}^{(\bar{Y}_k)}\Big|$ by ${\sigma}_{j}^{(Y_k)}$ , $\Bar{\mu_{j}}^{(\bar{Z}_k)}$ by $\mu_{j}^{(Z_k)}$ , $\Bar{\mu_{l}}^{(\bar{Z}_k)}$ by $\mu_{l}^{(Z_k)}$, $ {\Sigma}_{j}^{(\bar{Z}_k)}$ by ${{\sigma}_{j}^{(Z_k)}}^2 $  and ${\Sigma}_{l}^{(\bar{Z}_k)}$ by ${{\sigma}_{l}^{(Z_k)}}^2$, where each vector is replaced by its scalar analogue. Again $\mu_{j}^{(Z_k)}=T\mu_{Z_{k}^{(2)}}+j$, ${{\sigma}_{j}^{(Z_{k})}}^2=T^2\sigma_{Z_{k}^{(2)}}^2$ and  ${{\sigma}_{j}^{(Y_{k})}}^2=T^2{\sigma_{X_k}}^2+\sigma_{Z_{k}^{(2)}}^2 $, $\forall j=0(1)R$. 

Therefore, for a fixed Gaussian quantum channel, the quantum received signal can be calculated as the convolution product of the Gaussian distributed transmitted signal and the quantum noise signal.
Considering the hybrid quantum noise and the Gaussian transmitted signal model as discussed above, the capacity of the quantum channel can be expressed as follows 
\begin{equation}
    \begin{split}
       C_{k} &
          =\sum_{j=0}^{R}\frac{e^{-\lambda}\lambda^j}{j!}\Bigg(-\log_{2}\Big(\frac{e^{-\lambda}\lambda^j}{j!}\Big)
         + \frac{1}{2}\log_2{\Big(2\pi e \big({T^2\sigma_{X_k}}^2+\sigma_{Z_{k}^{(2)}}^2\big)\Big)}\\
         & + {\log_{2}}\Big(\sum_{l=0}^{R}\frac{e^{-\lambda}\lambda^j}{j!}\mathcal{N} \Big(\mu_{Z_{k}^{(2)}}+i;\mu_{Z_{k}^{(2)}}+l,\,2\sigma_{Z_{k}^{(2)}}^2\Big)\Big)\Bigg)\\
         & =\sum_{j=0}^{R}\frac{e^{-\lambda}\lambda^j}{j!}\Bigg(-\log_{2}\Big(\frac{e^{-\lambda}\lambda^j}{j!}\Big)
         + \frac{1}{2}\log_2{\Big(2\pi e \big(T^2{\sigma_{X_k}}^2+\sigma_{Z_{k}^{(2)}}^2\big)\Big)}\\
         & +\log_{2}\Big(\sum_{l=0}^{R}\frac{e^{-\lambda}\lambda^l}{l!}.\frac{1}{\sqrt{2}\sigma_{Z^{(2)}_{k}}\sqrt{2\pi}}e^{-\frac{1}{2}\Big[\frac{j-l}{\sqrt{2}\sigma_{Z^{(2)}_{k}}}\Big]^2}\Big)\Bigg),
      \label{eq the channel capacity }
   \end{split}
\end{equation}  
where 
\begin{equation}
\mathcal{N} \Big(\mu_{Z_{k}^{(2)}}+j;\mu_{Z_{k}^{(2)}}+l,\,2{\sigma_{Z_{k}^{(2)}}}^2\Big)=\frac{1}{\sqrt{2}\sigma_{Z^{(2)}_{k}}\sqrt{2\pi}}e^{-\frac{1}{2}\Big(\frac{\mu_{Z^{(2)}_{k}}+j-l-\mu_{Z^{(2)}_{k}}}{\sqrt{2}\sigma_{Z^{(2)}_{k}}}\Big)^2}. 
\end{equation}


\subsection{Secret Quantum Key Rate Estimation for \ac{FSO} Satellite Quantum Channel}
\vspace{0.15cm}

Revisiting the asymptotic \ac{SKR} derivation from the non-asymptotic perspective is also beneficial. According to the asymptotic perspective, the secret keys remain secure against general attacks, even in the finite-size regime. The determination of the asymptotic \ac{SKR} given by \cite{dequal2021_SAT_CV_QKD},

\begin{equation}
    \begin{split}
        & K=\lim_{N \rightarrow \infty}{}\Big\{\frac{1}{N}\Big(H(Y^{(N)})-leak^{(N)}_{EC})\Big)-\Psi(\Gamma^{(N)})\Big\}
        =\beta I_{AB}-\chi_{BE},
        \label{eq key rate}
    \end{split}
\end{equation}
 where
 \begin{equation}
     \beta I_{AB} =\lim_{N \rightarrow \infty}{}\Big\{\frac{1}{N}\Big(H(Y^{(N)})-leak^{(N)}_{EC})\Big)\Big\}, \,
  \chi_{BE}=\lim_{N \rightarrow \infty}{}\Psi(\Gamma^{(N)}).
    \label{eq part2}
 \end{equation}
Here $H(Y^{(N)})$ represents the empirical entropy of the sequence $Y^{(N)}$, with the superscript $N$ explicitly indicating the dependence of these variables on the block length, the term $leak_{EC}^{(N)}$ denotes the number of bits disclosed during the error correction process, wherein Bob provides auxiliary information to Alice, facilitating her accurate prediction of $Y$. Furthermore, $\Psi(\Gamma^{(N)})$ measures Eve's accessible information denoted by $\chi_{BE}$, which will be elucidated upon subsequently. This detail underlines our focus on the asymptotic behavior of the \ac{SKR}, allowing us to overlook discretization effects for this analysis. 
The utility of \eqref{eq key rate} lies in its guidance on calculating $\beta I_{AB}$ and $\chi_{BE}$ as per the Devetak-Winter bound. This equation simplifies determining these critical parameters, essential for evaluating the efficiency and security of \ac{QKD} systems in the context of quantum cryptography \cite{PirandolasatQComm2021}.

While the calculations of $\beta I_{AB}$ and $\Psi(\Gamma)$ are relatively straightforward for a fixed Gaussian channel with consistent transmission efficiency and excess noise \cite{acosta2024sat_qkd,lodewyck2007quantum}, they become more complex for fading quantum channels, leading to varied findings in existing work.  Specifically, employing Gaussian modulation allows for approaching the channel capacity, factoring in a reconciliation efficiency factor $\beta$, leading to \cite{lapidoth2002fading_channels_sum}

\begin{equation}
     \beta I_{AB} =\beta \frac{1}{N}\sum_{k=1}^{N}C_{k},
     \label{info_AB}
\end{equation}
where 
$C_k$ is given by \eqref{eq the channel capacity } $\forall k=1(1)N$.

Apart from the generic hybrid quantum noise, the free-space satellite channel suffers from associated noise referred to as excess noise, denoted by $\epsilon$. Furthermore, we account for imperfections in Bob's detectors through two specific parameters, e.g., the detection efficiency $\eta$ and the electronic noise $\nu_{ele}$. The quantum channel is characterized by its transmission coefficient $T \leq 1$ and excess noise $\epsilon$.  The total added noise referred to the quantum channel's input due to losses $1/T-1$, and the excess noise $\epsilon$ (in shot noise units ) is given by $\chi_{line}=1/T-1+\epsilon$ \cite{lodewyck2007quantum}. The signal then reaches Bob's detector, which is modeled by assuming that the signal is further attenuated by a factor $\eta$, the detection losses mixed with some thermal noise, specifically electronic noise $\vartheta_{ele}$ added by the detection electronics, expressed in shot noise units. The noise introduced by the realistic homodyne detector is $\chi_{hom} = (1+\vartheta_{ele})/\eta)-1$ \cite{lodewyck2007quantum}. The total cumulative noise added between Alice and Bob specifically for the free-space quantum satellite channel is given by \(\chi_{tot}=\chi_{line}+\chi_{hom} /T\), referred to the channel input keys \cite{lodewyck2007quantum}.

Now, focusing on the component $\chi_{BE}$,  the term $\Psi(\Gamma^{(N)})$ assesses the information Eve could potentially access. Specifically, $\Gamma^{(N)}$ provides a conservative estimation of the (average) covariance matrix associated with the state that Alice and Bob would theoretically share in an entanglement-based rendition of their protocol. The function $\Psi$ is characterized as \cite{lodewyck2007quantum}

\begin{equation}
\Psi(\Gamma)=S(\boldsymbol{\rho}_{AB})-S(\boldsymbol{\rho}_{BE}),
    \label{eq}
\end{equation}
where $S(\boldsymbol{\rho})$ is defined as the von Neumann entropy of the quantum state $\boldsymbol{\rho}$. For an $n$-mode Gaussian state $\rho$, this entropy \cite{acosta2024sat_qkd} is expressed as 
\(S(\boldsymbol{\rho}) =\sum_{j}^{}G\Big(\frac{\lambda_{j}-1}{2}\Big),\)
where \(G(x) = (x + 1)\log_2(x + 1)-x \log_2 x,\) and $\lambda_{j}$ are the symplectic eigenvalues \cite{son2022Symplectic_eigenvalues} of the covariance matrix $\boldsymbol{\gamma}$ characterizing $\boldsymbol{\rho}$ . 
For the sake of simplicity, let us consider the average variance of all transmitted signal 
\begin{equation}
     \sigma_{X} =\frac{1}{2N}\sum_{k=1}^{2N}\sigma_{X_k},
     \label{}
\end{equation}
where 
$\sigma_{X_k}$ is given by \eqref{eq The p.d.f. of transmitted signal} $\forall k=1(1)2N$.

The entropy $S(\boldsymbol{\rho}_{AB})$ is calculated from the symplectic eigenvalues $\lambda_{1}, \lambda_{2}$ of the covariance 
matrix  $\boldsymbol{\gamma}_{AB}$=
$\begin{bmatrix}
 \gamma_{A} & \sigma_{AB}\\
 \sigma_{AB}^t &  \gamma_{B}
\end{bmatrix}$, and the symplectic eigenvalues of $\boldsymbol{\gamma}_{AB}$ are given by 
\begin{equation}
    \lambda_{1,2}^2= \frac{1}{2}\Bigl[A \pm \sqrt{A^2-4B}\Bigl],
    \label{lambda1_2}
\end{equation}
where \(A=\sigma_{X}^2(1-2T)+2T+T^2(\sigma_{X}^2+\chi_{line})^2,\) and \(B=T^2(\sigma_{X}^2\chi_{line}+1)^2 .\) Similarly, the entropy $S(\boldsymbol{\rho}_{BE})$ is determined from the symplectic eigenvalues $\lambda_{3},\lambda_{4},\lambda_{5}$ of the covariance matrix characterizing the state $\boldsymbol{\rho}_{E}^{Y_B}$ of Eve's system conditional on Bob's measurement outcome $Y_B$. The symplectic eigenvalues $\lambda_3,\lambda_4$ are given by
\begin{equation}
    \lambda_{3,4}^2= \frac{1}{2}\Bigl[C \pm \sqrt{C^2-4D}\Bigl],
    \label{lambda3_4}
\end{equation}
where 
\begin{equation}
    C =\frac{\sigma_{X_k}^2\sqrt{B} + T(\sigma_{X}^2 + \chi_{line}) + A\chi_{hom}}{T(\sigma_{X}^2 + \chi_{tot})}, \,
    D = \sqrt{B}\frac{\sigma_{X}^2 + \sqrt{B}\chi_{hom}}{T(\sigma_{X}^2 + \chi_{tot})},\label{D}
\end{equation} and
the last symplectic eigenvalue is simply $\lambda_5 = 1$. The Holevo information bound is given by \cite{lodewyck2007quantum,ghalaii2022quantumSatcomm}
\begin{equation}
        \chi_{BE}=G\Bigg(\frac{\lambda_{1}-1}{2}\Bigg)+G\Bigg(\frac{\lambda_{2}-1}{2}\Bigg)-G\Bigg(\frac{\lambda_{3}-1}{2}\Bigg)-G\Bigg(\frac{\lambda_{4}-1}{2}\Bigg),
        \label{info_BE}
\end{equation}
and from \eqref{eq key rate}, \eqref{info_AB} and \eqref{info_BE} the \ac{SKR} is given as 
\begin{multline}
        K= \beta I_{AB}-\chi_{BE}\\ =\beta \frac{1}{N}\sum_{k=1}^{N}C_{k}-\Bigl[ G\Big(\frac{\lambda_{1}-1}{2}\Big)+G\Big(\frac{\lambda_{2}-1}{2}\Big)-G\Big(\frac{\lambda_{3}-1}{2}\Big)-G\Big(\frac{\lambda_{4}-1}{2}\Big)\Bigl].
\end{multline}

This work scrutinizes practical \ac{QKD} rates by introducing a hybrid quantum channel model tailored for free-space \ac{QKD} by elaborating on the foundational theoretical limitations of quantum communication. This model is tailored to the fading properties present in satellite-to-ground links while considering the satellite's dynamic orbit. Our security analysis includes considering finite-size key distribution and composable security, which are crucial for practical implementations. This paper expands the theoretical foundation and practical application of quantum communications via satellite links, extending existing analyses from ground-based free-space settings to the more complex scenario of ground-to-satellite communications. This transition involves longer travel distances of optical signals through the atmosphere, which must account for variable altitudes and zenith angles, whether in uplink or downlink configurations. Crucially, this environment includes additional atmospheric disturbances causing noises such as refraction, extinction, and turbulence, alongside varied sources of background noise like planetary albedos and sky brightness. These elements are crucial in modeling the conditions under which quantum communication occurs, offering a detailed understanding of the potential for secure key generation and entanglement distribution using satellites, regardless of the time of day or link direction. It demonstrates that \ac{QKD} via \ac{CV} systems is feasible and highly effective for downlink and uplink configurations operating during daytime and nighttime. Furthermore, it highlights the superiority of satellite-based quantum key distribution over traditional existing networks.

\section{Results and Numerical Analysis}

\begin{figure}%
    \centering
    \subfloat[\centering ]{{\includegraphics[width=7.5cm]{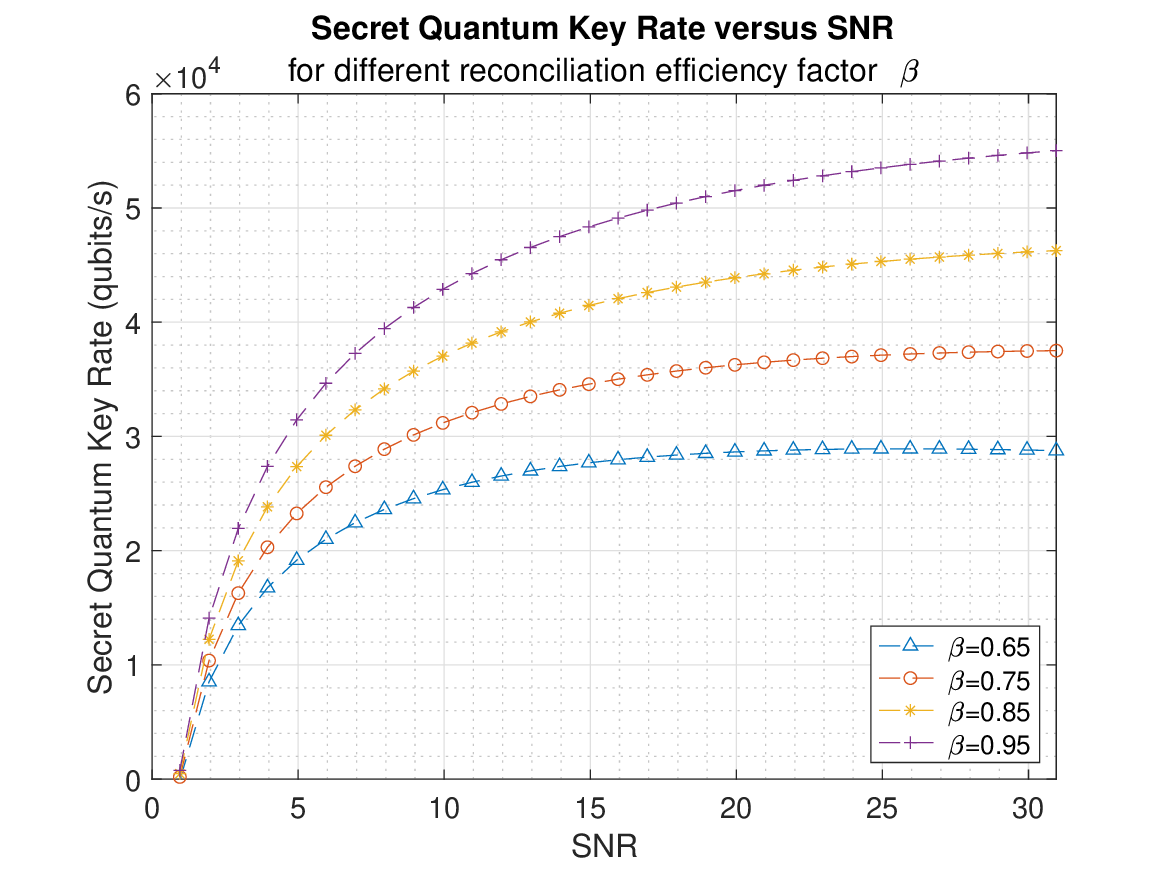} }}%
    \qquad
    \subfloat[\centering ]{{\includegraphics[width=7.5cm]{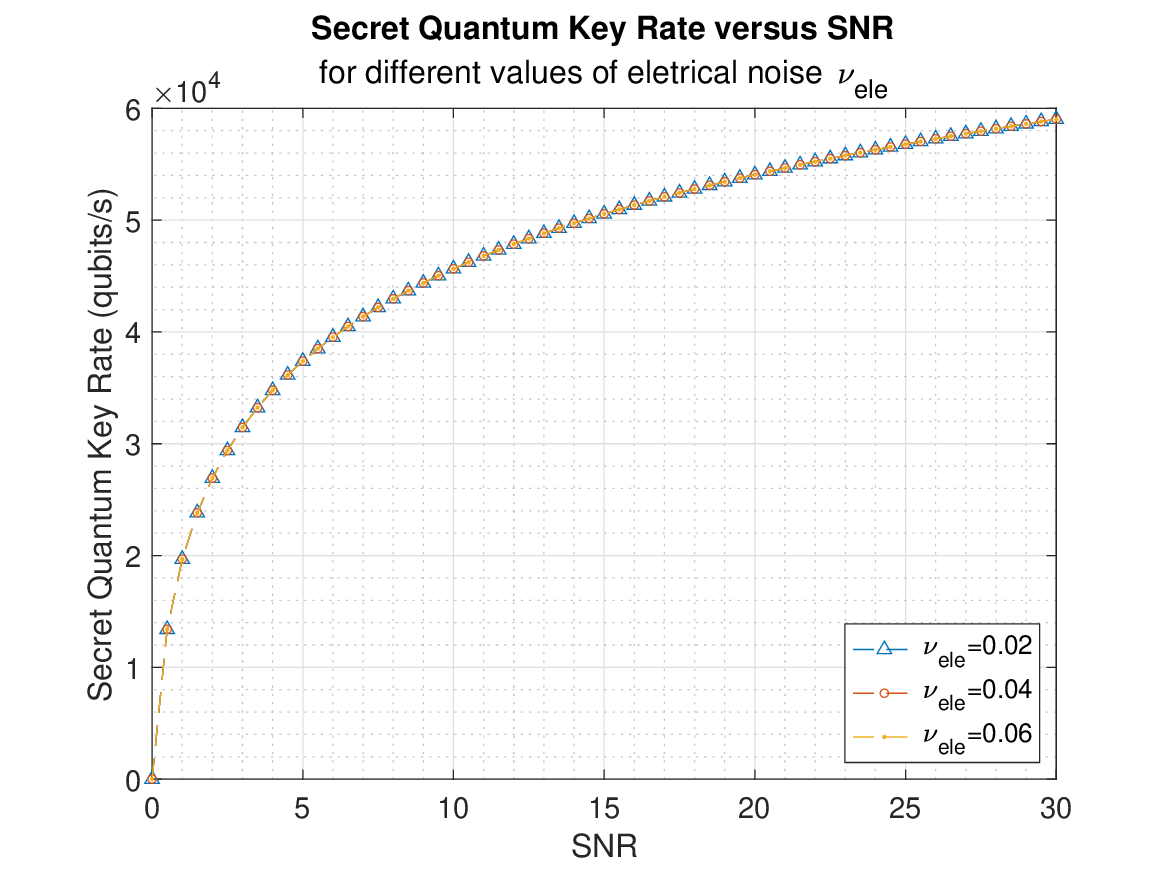} }}%
    \caption{(a) Impact of Reconciliation Efficiency on Quantum SKR
    , (b) Impact of Electrical Noise on Quantum SKR.}%
    \label{fig reconcilation efficiency and electrical noise}%
\end{figure}

In this section, we delve into the analysis of Quantum \ac{SKR} in Free-Space Atmospheric Satellite Quantum Channels, focusing on its relationship with the \ac{SNR}. The study explores how various channel parameters, including reconciliation efficiency, electrical noise, transmission coefficient, transmission efficiency, excess noise, detection efficiency, and quantum Poissonian noise, influence the SKR in the context of free-space atmospheric satellite quantum channels. The findings offer a comprehensive understanding of the factors affecting the performance of \ac{QKD} in satellite-based systems.


Reconciliation efficiency \(\beta\) is a crucial factor impacting the quantum \ac{SKR}. This efficiency measures the effectiveness of error correction between the sender (Alice) and the receiver (Bob). Fig. \ref{fig reconcilation efficiency and electrical noise} (a) illustrates the dependence of quantum \ac{SKR} on different \(\beta\) values from $0.65, 0.75,0.85,$ and $0.95$, keeping other parametric values fixed with $\lambda =2$, $\epsilon =0.005$ shot noise units, $T=1$, $\eta=0.606$, and $\nu_{ele}=0.041$ shot noise units. The sensitivity of \ac{SKR} to \(\beta\) is significant because it directly affects the amount of mutual information that can be converted into a secure key. A higher \(\beta\) value indicates more efficient reconciliation, leading to a higher SKR, as fewer bits are lost during error correction. Conversely, a lower \(\beta\) reduces the \ac{SKR}, even under ideal channel conditions, because a larger portion of shared information must be discarded.  The results underscore the importance of optimizing \(\beta\) to maximize \ac{SKR} in satellite-based quantum channels.


Electrical noise \(\nu_{ele}\) represents the classical noise introduced by detection and electronic processing systems at the receiver's end.  Fig. \ref{fig reconcilation efficiency and electrical noise} (b) illustrates the dependence of quantum \ac{SKR} on different \(\nu_{ele}\) values ranging from $0.020,0.040$ and $0.060$ shot noise units, with $\lambda =2$, $\epsilon=0.005$ shot noise units, $\beta=0.95$, $\eta=0.606$, and $T=1$.  The variations in \(\nu_{ele}\) within its permissible range do not significantly impact the \ac{SKR}. This indicates that electrical noise is not a primary limiting factor for \ac{SKR} in these channels.  The analysis reveals that within its permissible range, \(\nu_{ele}\) has a minimal effect on the \ac{SKR}. This suggests that the system is designed to function effectively even with typical levels of electrical noise encountered in free-space atmospheric satellite channels.   The robustness of \ac{QKD} protocols to moderate levels of classical noise due to error correction and privacy amplification steps explains this insensitivity for the satellite-based quantum channels. However, if \(\nu_{ele}\) exceeds the permissible range, it could degrade the \ac{SKR} by increasing the error rate beyond what can be corrected, thereby reducing the amount of secure key generated. This highlights the importance of maintaining low electrical noise levels while emphasizing the system's resilience within the tested parameters.

\begin{figure}%
    \centering
    \subfloat[\centering ]{{\includegraphics[width=7.5cm]{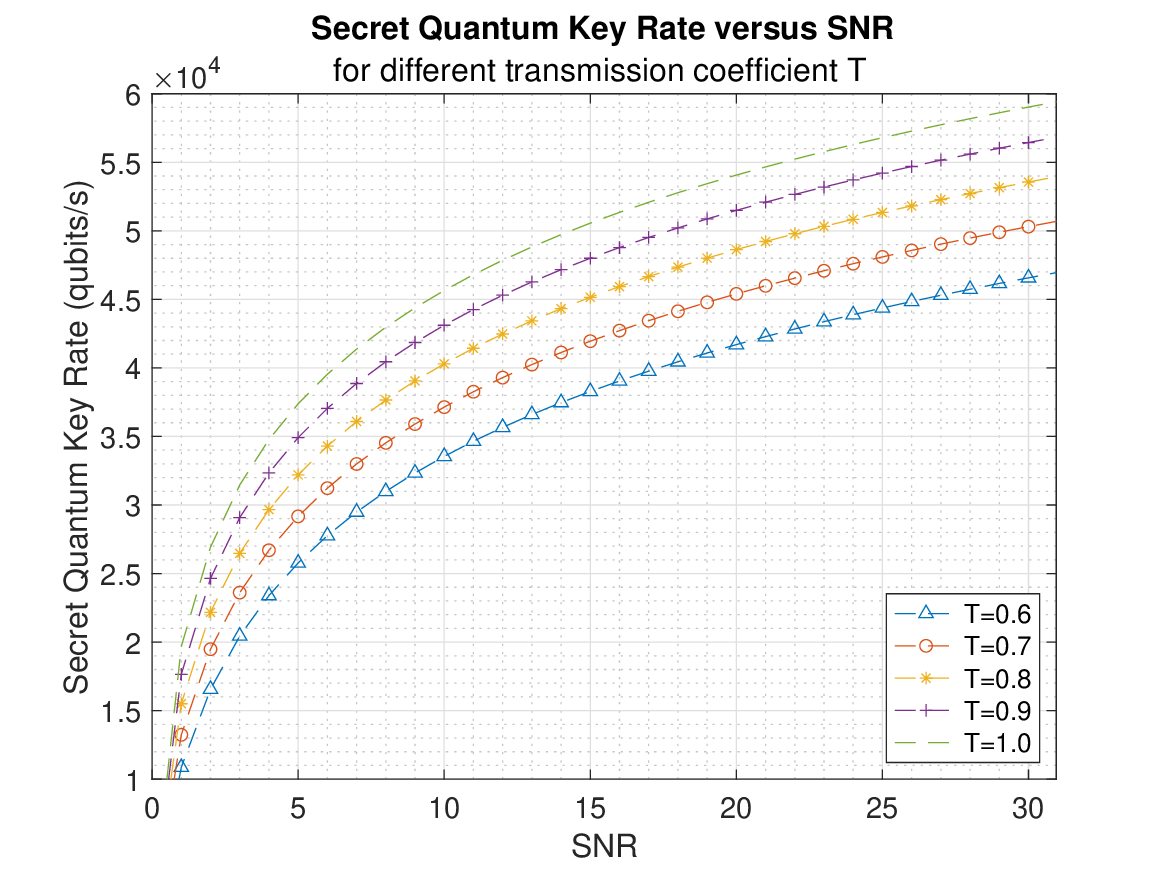} }}%
    \qquad
    \subfloat[\centering ]{{\includegraphics[width=7.5cm]{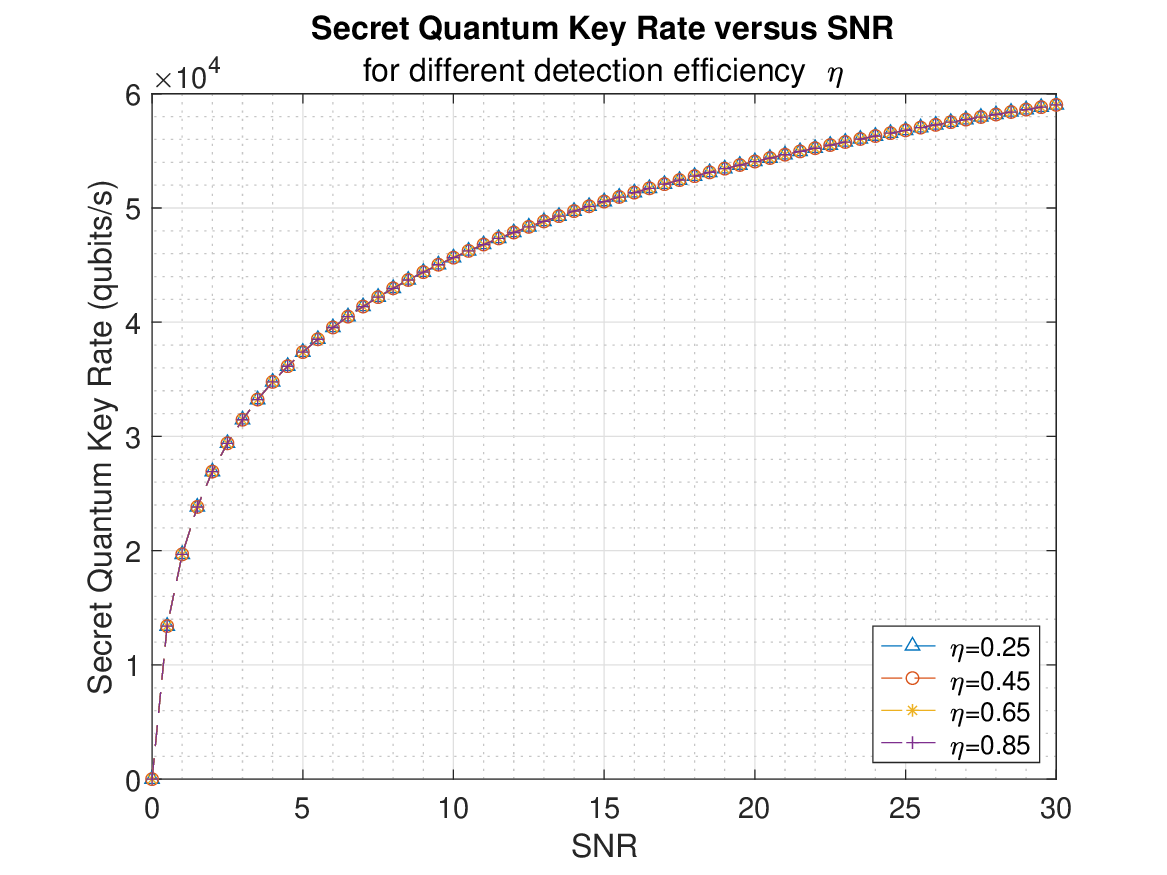} }}%
    \caption{(a) Impact of Transmission Coefficient on Quantum SKR, (b) Impact of Detection Efficiency on Quantum SKR.}%
    \label{fig transmissionCoefficient and detection_efficiency}%
\end{figure}


The transmission coefficient \(T\) is a vital parameter influencing the quantum \ac{SKR}. \(T\) reflects the fraction of quantum states successfully transmitted from Alice to Bob through the terrestrial quantum channel. A higher transmission coefficient improves \ac{SKR} by ensuring that more quantum states reach the receiver. The \ac{SKR}'s sensitivity to \(T\) is substantial because it encapsulates the effects of absorption, scattering, and other losses during quantum state transmission. In free-space quantum communication, particularly in satellite channels, atmospheric conditions and distance significantly impact \(T\). Fig. \ref{fig transmissionCoefficient and detection_efficiency} (a) illustrates the dependence of quantum \ac{SKR} on different \(T\) values ranging from $T=0.6, 0.7, 0.8, 0.9$ up to $1$ with setting other parametric values as $\lambda =2$, $\epsilon=0.005$ shot noise units, $\beta=0.95$, $\eta=0.606$  and $\nu_{ele}=0.041$ shot noise units.  It shows that the higher transmission coefficients lead to better \ac{SKR} performance, as fewer quantum states are lost in transit. Optimizing \(T\) is, therefore, critical for enhancing \ac{SKR}, especially in satellite-based \ac{QKD} systems where atmospheric conditions and distance can greatly influence transmission efficiency.


Detection efficiency \(\eta\) is crucial for accurately measuring quantum states. It determines the fraction of incoming quantum states that are successfully detected and measured. Fig. \ref{fig transmissionCoefficient and detection_efficiency} (b) illustrates the dependence of quantum \ac{SKR} on different \(\eta\) values ranging from $0.25,0.45$, $0.65$ and $0.85$, with $\lambda =2$, $\epsilon=0.005$ shot noise units, $\beta=0.95$, $\nu_{ele}=0.041$ shot noise units, and $T=1$. The \ac{SKR}'s insensitivity to \(\eta\) in the observed range implies that the detection mechanism in the system is robust and reliable. Within the studied range, most transmitted states that reach the receiver are successfully detected, leaving the \ac{SKR} largely unaffected by variations in \(\eta\). However, if \(\eta\) were significantly lower, it would decrease the detected signal, consequently reducing the \ac{SKR}. This work illustrates that changes in \(\eta\) do not substantially impact the \ac{SKR}, indicating the system's reliability under varying conditions.

\begin{figure}%
    \centering
    \subfloat[\centering ]{{\includegraphics[width=7.5cm]{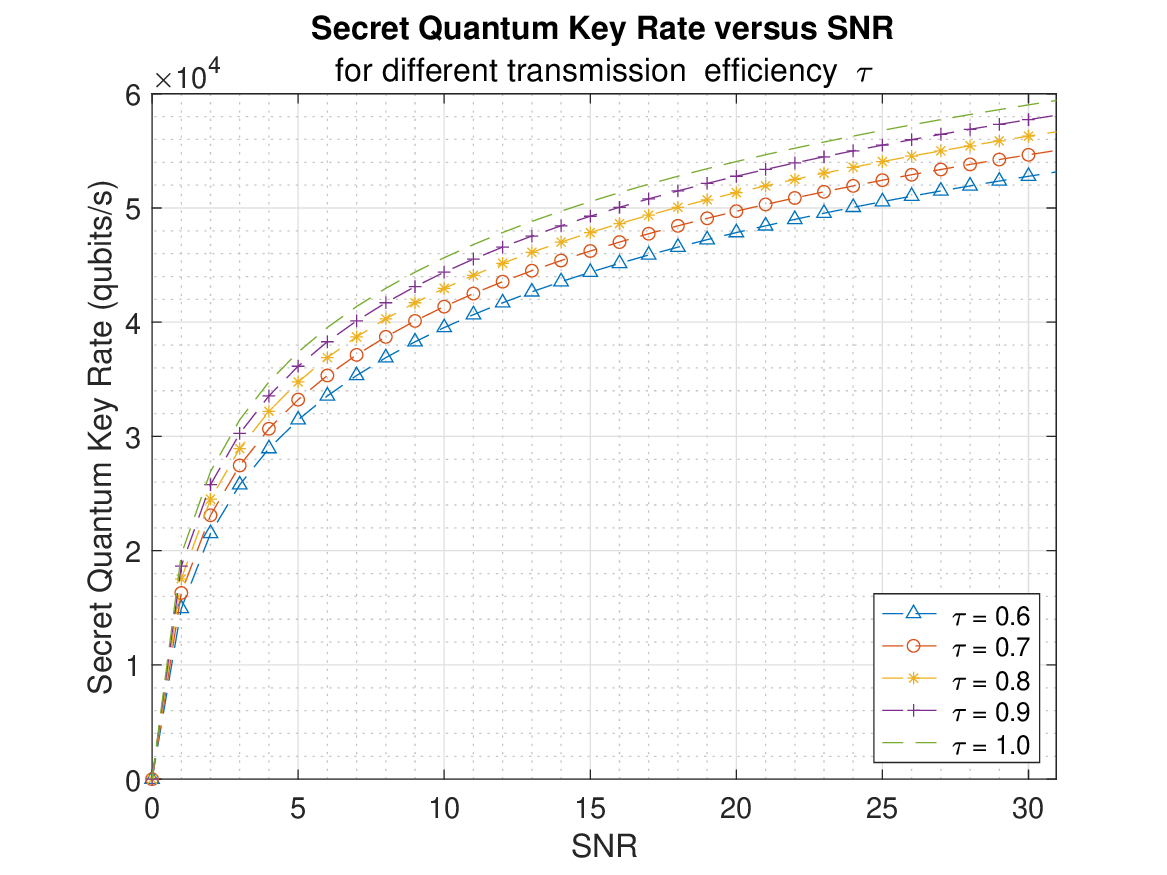} }}%
    \qquad
    \subfloat[\centering ]{{\includegraphics[width=7.5cm]{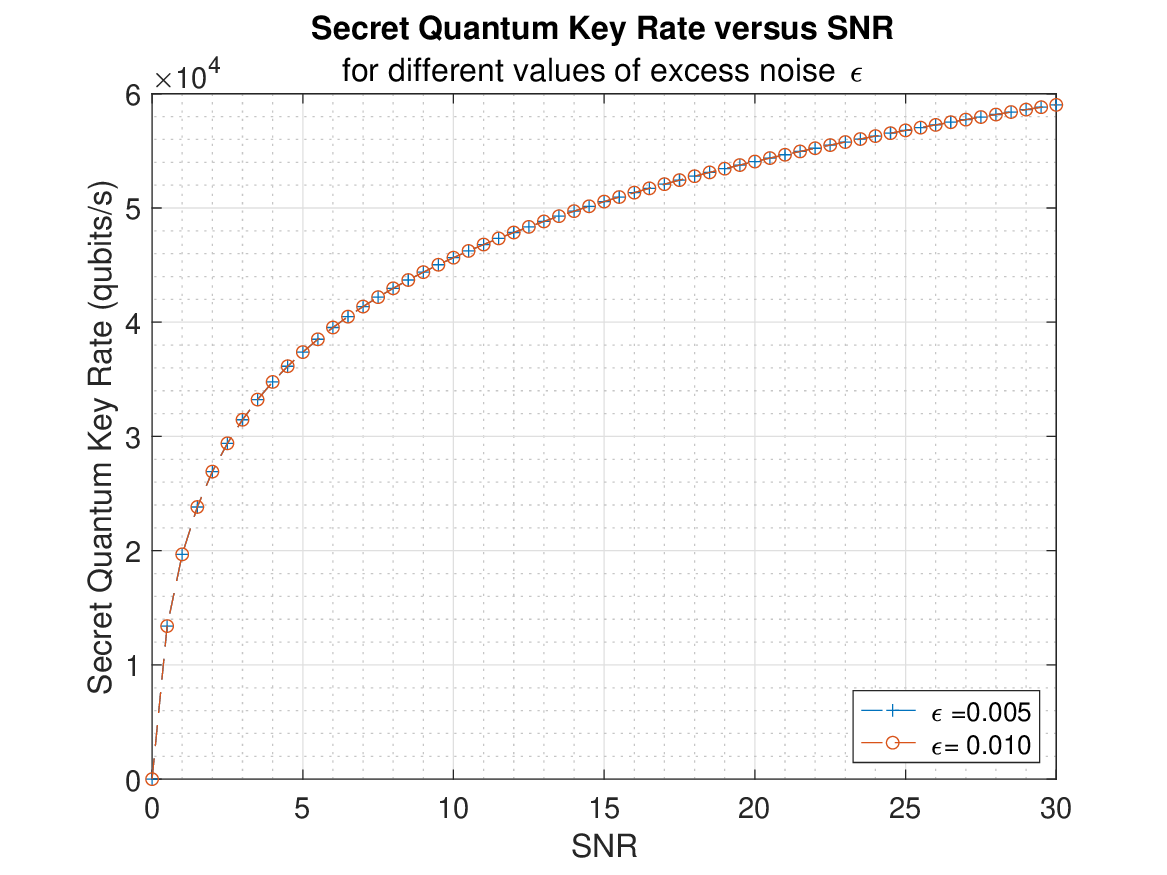} }}%
    \caption{(a) Impact of Transmission Efficiency on Quantum SKR,  (b) Impact of Excess Noise on Quantum SKR.}%
    \label{fig transmissionEF and excess noise}%
\end{figure}


Transmission efficiency \(\tau\) is another critical factor affecting the quantum \ac{SKR}. This parameter measures how effectively quantum states are transmitted through the channel without being lost due to absorption, scattering, or diffraction. Fig.\ref{fig transmissionEF and excess noise} (a) demonstrates the variation of \ac{SKR} with different \(\tau\) values ranging from $\tau=0.6, 0.7, 0.8, 0.9$ up to $1$, and $T=\sqrt{\tau}$ with setting other parametric values as $\lambda =2$, $\epsilon=0.005$ shot noise units, $\beta=0.95$, $\eta=0.606$  and $\nu_{ele}=0.041$ shot noise units. The \ac{SKR} is highly sensitive to \(\tau\) because higher transmission efficiency means more quantum states reach the receiver, increasing the likelihood of successful \ac{QKD} and, consequently, a higher \ac{SKR}. Conversely, lower \(\tau\) results in more quantum states being lost in transit, reducing the usable data for key generation and decreasing the \ac{SKR}. It emphasises the importance of optimizing \(\tau\) to enhance \ac{SKR} in \ac{FSO} satellite-based quantum channels.


Excess noise \(\epsilon\) represents the noise in the quantum channel beyond what is expected from quantum mechanical limits. Fig. \ref{fig transmissionEF and excess noise} (b) demonstrates the variation of \ac{SKR} with different \(\epsilon\) values  $0.005$ up to $0.010$ shot noise units, with setting other parametric values as $\lambda =2$, $\beta=0.95$, $\eta=0.606$, $T=1$  and $\nu_{ele}=0.041$ shot noise units. The \ac{SKR} shows minimal sensitivity to \(\epsilon\) within its permissible range, indicating that modern \ac{QKD} protocols are robust against reasonable noise levels. However, if \(\epsilon\) exceeds certain thresholds, it would introduce additional errors that could not be reconciled, reducing the \ac{SKR}. This work shows that excess noise does not significantly impact the \ac{SKR} within the permissible range. This suggests that the system is well-designed to tolerate typical environmental and instrumental noise levels without significant degradation in performance.

\begin{figure}%
    \centering
    \subfloat[\centering ]{{\includegraphics[width=7.5cm]{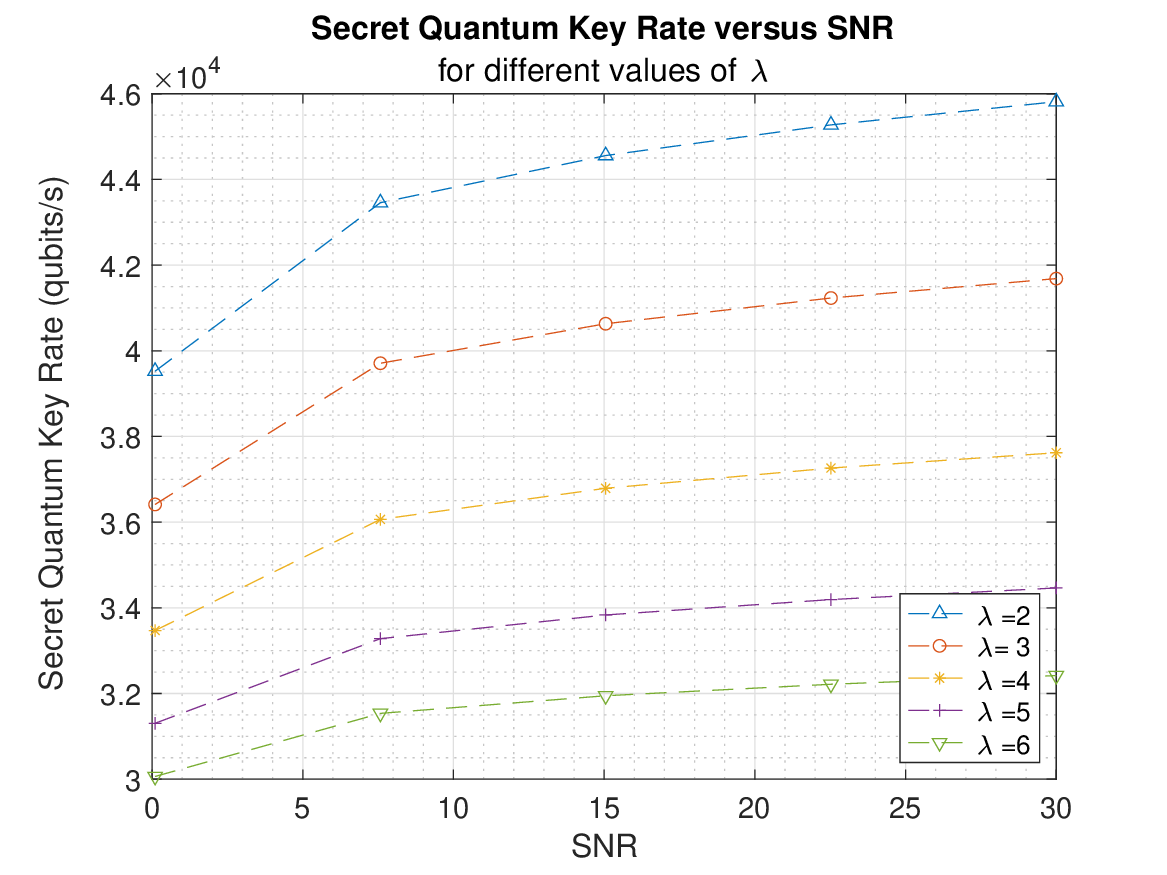} }}%
    \qquad
    \subfloat[\centering ]{{\includegraphics[width=7.5cm]{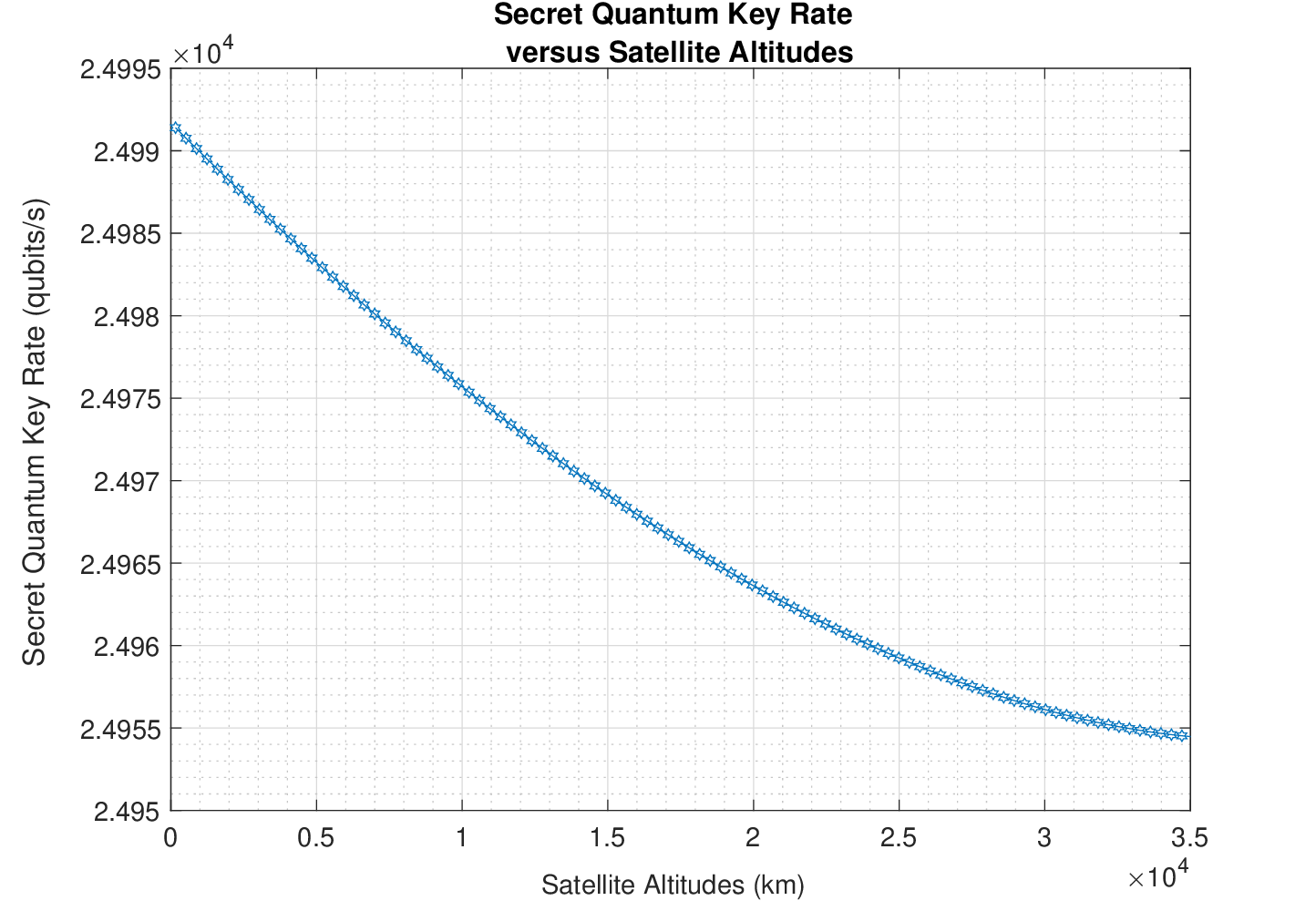} }}%
    \caption{(a) Impact of Quantum Poissonian Noise on Quantum SKR, (b) Impact of Satellite altitudes on Quantum SKR.}%
    \label{fig lambda_keyrate and distance}%
\end{figure}


Quantum Poissonian noise \(\lambda\) significantly influences the quantum \ac{SKR}. This parameter represents the level of quantum Poissonian noise in the channel, originating from the random nature of photon emission in quantum communication systems. Fig. \ref{fig lambda_keyrate and distance} (a) illustrates the dependence of quantum \ac{SKR} on different \(\lambda\) values ranging from $2,3,4,5$ and $6$, with $\epsilon=0.005$ shot noise units, $\beta=0.95$, $\nu_{ele}=0.041$ shot noise units, $\eta=0.606$, and $T=1$. A larger \(\lambda\) increases the noise photons present, raising the error rate and reducing the \ac{SKR}, as more signals may be confused with noise rather than legitimate quantum states. Conversely, a smaller \(\lambda\) implies fewer noise photons, improving the \ac{SKR} by increasing the \ac{SNR} and making distinguishing between quantum states carrying key information and noise easier.  This work demonstrates how different \(\lambda\) values can substantially affect the \ac{SKR}. Optimizing \(\lambda\) is essential for achieving high \ac{SKR} in free-space atmospheric satellite channels.


Finally, the analysis examines how the quantum \ac{SKR} varies with satellite altitudes $h$. Fig. \ref{fig lambda_keyrate and distance} (b) shows that the quantum \ac{SKR} decreases as satellite altitudes increase considering transmission efficiency \(\tau\) as a function of satellite altitudes and \ac{SKR} as a function of transmission efficiency, setting the primary parametric values as $\epsilon=0.005$ shot noise units, $\beta=0.95$, $\nu_{ele}=0.041$ shot noise units, $\eta=0.606$, and $\lambda=2$.  As satellite altitudes increase, the transmission efficiency \(\tau\) decreases due to factors like atmospheric attenuation, beam divergence, and other losses over high altitudes. This reduction in \(\tau\) leads to a lower \ac{SKR} as fewer quantum states reach the receiver. For \ac{CV} satellite quantum channel, the findings underscore the importance of optimizing transmission efficiency and noise management strategies to ensure effective quantum communication over extended altitudes, where losses are more pronounced. 


This work comprehensively analyzes the factors influencing the quantum \ac{SKR} in free-space atmospheric satellite quantum channels. The study highlights the significant impact of reconciliation efficiency \(\beta\), transmission efficiency \(\tau\), and quantum Poissonian noise \(\lambda\) on \ac{SKR}. In contrast, electrical noise and excess noise show minimal impact within their permissible ranges. The results emphasize optimizing these parameters, particularly as satellite altitudes increase, to maintain high \ac{SKR} levels. This investigation lays the groundwork for future work to enhance the performance of satellite-based quantum communication systems, ensuring the secure and efficient transmission of quantum information across global distances.

\section{Conclusion}


This work advances the application of \ac{QKD} in free-space atmospheric satellite-based quantum communication. We developed a novel free-space atmospheric satellite quantum channel model and investigated the main parameters influencing such channels' quantum \ac{SKR}. Unlike existing models that consider noise in quantum channels as purely Gaussian distributed, our model incorporates a hybrid noise framework that includes both quantum Poissonian noise and classical \ac{AWGN}. This approach acknowledges the dual vulnerabilities of quantum channels to quantum and classical noise, providing a more realistic assessment of the \ac{SKR}. The paper identifies several critical parameters significantly impacting the \ac{SKR} in free-space atmospheric satellite quantum channels. These parameters include reconciliation efficiency (\(\beta\)), transmission coefficient (\(T\)), transmission efficiency (\(\tau\)), and the quantum Poissonian noise parameter (\(\lambda\)). Our findings highlight that optimizing these parameters is essential for maximizing the \ac{SKR} and ensuring robust quantum communication. Furthermore, the work concludes that the quantum \ac{SKR} decreases with increasing satellite altitude. This declination is attributed to increased losses and reduced signal strength over longer distances, underscoring the importance of maintaining high transmission efficiency and effective noise management strategies in satellite-based \ac{QKD} systems. This paper contributes to the field by proposing a more realistic quantum channel model and identifying the influential factors affecting the \ac{SKR} in free-space atmospheric satellite quantum communication. The insights gained from this paper lay the groundwork for further work to enhance the performance and security of satellite-based quantum communication systems, facilitating the global transmission of quantum information.

\begin{acronym}
    \acro{AWGN}{Additive-White-Gaussian Noise}
    \acro{SKR}{Secret Key Rate}
    \acro{QKD}{Quantum Key Distribution}
    \acro{PNS}{Photon Number Splitting}
    \acro{CV-QKD}{Continuous-Variable QKD}
    \acro{FSO}{Free-Space Optics}
    \acro{MDI}{Measure-device-independent}
    \acro{DV-QKD}{Discrete-Variable QKD}
    \acro{CV}{continuous variables}
    \acro{DV}{discrete variables}
    \acro{CPTP}{completely positive, trace preserving}
    \acro{PM}{Prepare-and-Measure}
    \acro{PDTC}{Probability Distribution of Transmission Coefficient}
    \acro{S/C}{spacecraft}
    \acro{p.d.f.}{probability density function}
    \acro{p.m.f.}{probability mass function}
    \acro{SNR}{Signal-to-Noise Ratio}
    \acro{SPS}{Single-Source Photon}
    \acro{GMM}{Gaussian Mixture Model}
    \acro{GMs}{Gaussian Mixtures}
    \acro{GS}{Ground Station}
\end{acronym}
\bibliographystyle{IEEEtran}
\bibliography{IEEEabrv,paper}

\end{document}